\documentclass[11pt]{article}

\usepackage[a4paper, margin=1.2in]{geometry}
\usepackage[utf8]{inputenc}
\usepackage[T1]{fontenc}
\usepackage[english]{babel}
\usepackage[authoryear, round]{natbib}
\usepackage{authblk}
\usepackage{amsmath,amsfonts,amssymb}
\usepackage{graphicx}
\usepackage[colorlinks=true, allcolors=blue]{hyperref}
\usepackage{braket}
\usepackage{mathtools}
\usepackage[dvipsnames]{xcolor}
\usepackage{tcolorbox}
\usepackage{enumerate}
\usepackage{booktabs}
\usepackage[framemethod=TikZ,xcolor=RGB]{mdframed}
\usepackage{bm}
\usepackage{multirow}
\usepackage{titlesec}

\titleformat{\section}{\normalfont\large\bfseries}{\thesection}{1em}{}
\titleformat{\subsection}{\normalfont\normalsize\bfseries}{\thesubsection}{1em}{}
\titleformat{\subsubsection}{\normalfont\small\bfseries}{\thesubsubsection}{1em}{}

\definecolor{mybrown}{RGB}{102,101,71}
\definecolor{myyellow}{RGB}{255,226,138}
\definecolor{mygreen}{RGB}{111,203,159}

\hypersetup{
    colorlinks=true,
    linkcolor=blue,
    filecolor=magenta,      
    urlcolor=blue,
    pdfpagemode=FullScreen,
    }

\newtheorem{proposition}{Proposition}
\newtheorem{definition}{Definition}[section]

\renewcommand{\a}{\boldsymbol{a}}
\renewcommand{\t}{\boldsymbol{t}}
\newcommand{\x}{\boldsymbol{x}}
\newcommand{\y}{\boldsymbol{y}}
\newcommand{\g}{\boldsymbol{g}}
\newcommand{\z}{\boldsymbol{z}}
\newcommand{\s}{\boldsymbol{s}}
\newcommand{\h}{\boldsymbol{h}}

\newcommand{\btheta}{\boldsymbol{\theta}}
\newcommand{\MMD}{\text{MMD}}
\newcommand{\MMDD}{\text{MMD}$^2$}

\renewcommand{\cite}{\citealp}

\renewcommand{\vec}{\boldsymbol}

\usepackage{tikzpagenodes}

\begin{document}

\title{Train on classical, deploy on quantum: scaling generative quantum machine learning to a thousand qubits}

\author[1,2,3]{Erik Recio-Armengol\thanks{erik.recio@icfo.eu}}
\author[1]{Shahnawaz Ahmed}
\author[1]{Joseph Bowles\thanks{joseph@xanadu.ai}}

\affil[1]{Xanadu, Toronto, ON, M5G 2C8, Canada}
\affil[2]{ICFO-Institut de Ciencies Fotoniques, The Barcelona Institute of Science and Technology, 08860, Castelldefels (Barcelona), Spain}
\affil[3]{Eurecat, Centre Tecnològic de Catalunya, Multimedia Technologies, 08005 Barcelona, Spain}

\date{}
    \maketitle
    \begin{abstract}
      We propose an approach to generative quantum machine learning that overcomes the fundamental scaling issues of variational quantum circuits. The core idea is to use a class of generative models based on instantaneous quantum polynomial circuits, which we show can be trained efficiently on classical hardware. Although training is classically efficient, sampling from these circuits is widely believed to be classically hard, and so computational advantages are possible when sampling from the trained model on quantum hardware. By combining our approach with a data-dependent parameter initialisation strategy, we do not encounter issues of barren plateaus and successfully circumvent the poor scaling of gradient estimation that plagues traditional approaches to quantum circuit optimisation. We investigate and evaluate our approach on a number of real and synthetic datasets, training models with up to one thousand qubits and hundreds of thousands of parameters. We find that the quantum models can successfully learn from high dimensional data, and perform surprisingly well compared to simple energy-based classical generative models trained with a similar amount of hyperparameter optimisation. Overall, our work demonstrates that a path to scalable quantum generative machine learning exists and can be investigated today at large scales. 
    \end{abstract}


\section{Introduction}
\subsection{The crisis of scalability in variational quantum machine learning}
The need for scalable models has always been at the forefront of machine learning research. Many breakthroughs, from the contrastive divergence algorithm (\cite{hinton2006reducing}), the use of backpropagation in neural networks (\cite{rumelhart1986learning,krizhevsky2012imagenet}), the removal of recurrence by self-attention in transformers (\cite{vaswani2017attention}), to the use of modern mixture of expert models (\cite{fedus2022switch,shazeer2017outrageously}) have been so impactful precisely because they enable the use of ever larger models on ever larger datasets. 
At the same time, the modern era of deep learning has continually reinforced the bitter lesson (\cite{sutton2019bitter}) that scaling to larger datasets and models is often the most fruitful path to improved performance. 

In contrast, most variational quantum machine learning algorithms face a seemingly fundamental barrier to scalability. By far the most well known of these is the phenomenon of barren plateaus (\cite{mcclean2018barren, cerezo2021cost, ragone2024lie,arrasmith2022equivalence,uvarov2021barren,fontana2024characterizing}), where gradients of randomly intialized circuits concentrate exponentially around zero as the number of qubits grows. Barren plateaus are however only the tip of the iceberg. Unlike gradients of classical neural networks---which can be computed at the same cost as the loss function---estimating gradients of quantum circuits is more costly than loss function estimation by a factor that scales with the number of circuit parameters\footnote{Note that some examples of circuit structures with favorable gradient estimation are known however (\cite{bowles2023backpropagation,chinzei2024trade,coyle2024training}).} (\cite{wierichs2022general,abbas2024quantum}). For circuits with thousands of parameters, this requires sampling from thousands of different circuits to obtain a single gradient vector. On top of this, the properties of quantum computers add additional burdens which translate to a large constant factor slow-down compared to classical neural networks. In particular, the sample rates of quantum computers are expected to be orders of magnitude slower than clock rates of modern CPUs and GPUs (\cite{hoefler2023disentangling}) and---unlike classical neural networks---quantum processing units are inherently stochastic, which means they need to be sampled from in order to estimate any values used for training. 

The situation is particularly concerning when one looks at the actual numbers involved: to perform a single epoch of gradient descent\footnote{Here we are assuming a loss function bounded in $[0,1]$ that we estimate using the parameter shift method to a precision $\epsilon\approx0.001$ using $1/\epsilon^2=10^6$ circuit samples.} on the MNIST handwritten digits dataset for a circuit with 10000 parameters (far below the overparameterisation threshold) would require collecting on the order of $10^{15}$ samples from the quantum computer, which translates to $38$ years of continual operation using a quantum computer operating at a MHz sampling rate. Given that models often also have to be trained many times in an initial hyperparameter search phase, and that the number of quantum computers will be (at least initially) limited, it seems inevitable that the standard approach of variational quantum machine learning will always be restricted to training small models on small datasets. 
In our opinion, this uncomfortable and often unrecognised reality thus severely limits the potential of variational quantum machine learning to deliver real societal value, and there is an urgent need to properly address the issue of scalability.

\begin{figure*}
    \includegraphics[width=\textwidth]{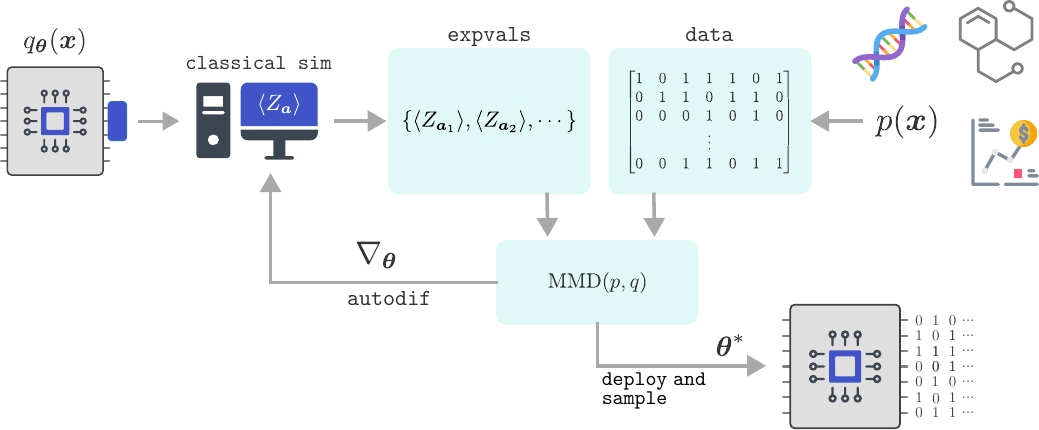}
    \caption{The method we use to train our quantum generative models. One first estimates a batch of expectation values $\{\langle Z_{\a_i}\rangle\}$ of Pauli Z words evaluated on the output distribution $q_{\btheta}(\x)$ of the quantum circuit. The class of circuits we use (parameterised IQP circuits), admit an efficient classical algorithm for this task, which is therefore performed on classical hardware. This information is combined with a dataset sampled from a ground truth distribution $p$ to provide an unbiased estimate of the squared maximum mean discrepancy between $p$ and $q_{\btheta}(\x)$. We use automatic differentiation to obtain estimates of gradients to train the circuit. Once the circuit is trained, the trained parameters $\btheta^*$ can be deployed on quantum hardware to generate samples. Since IQP circuits are believed to be hard to sample from classically, computational advantages are possible at this stage.}
    \label{fig:main}
\end{figure*}

\subsection{Overview of results: A new path to scalability}
\begin{figure*}
    \centering
    \includegraphics[scale=0.8]{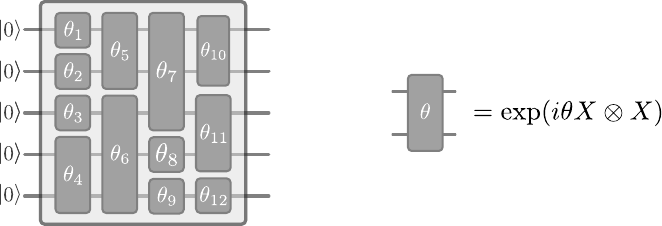}
    \caption{Parameterised IQP circuits consist of parameterised rotation gates whose generators are tensor products of Pauli-X operators.}
    \label{fig:piqpc}
\end{figure*}

In this work we theoretically and empirically demonstrate that, despite these barriers, a path to scalability exists, and can enable the use of large-scale circuits and datasets. Our approach (see Fig.~\ref{fig:main}) applies specifically to generative machine learning, where the quantum generative models correspond to parameterized instances of instantaneous quantum polynomial (IQP) circuits (Fig.~\ref{fig:piqpc}). The possibility to train such circuits at scale follows from the combination of two ingredients. First, \citet{rudolph2024trainability} have shown that the maximum mean discrepancy (MMD) loss function (\cite{JMLR:v13:gretton12a}), which can be used to train quantum generative models, can be efficiently cast as a classical mixture of expectation values of Pauli-Z words. Second, results by \citet{nest2010simulating} imply that expectation values of Pauli-Z words of IQP circuits can be estimated efficiently by a classical algorithm. Using the classical simulation algorithm of \citet{nest2010simulating} within the decomposition of the MMD loss function, one arrives at an efficient method to train parameterized IQP circuits with classical hardware alone. Crucially for generative learning, sampling from IQP circuits is expected to be hard\footnote{More specifically, inverse polynomial additive error sampling would imply collapse of the polynomial hierarchy to its second level, subject to either one of two additional assumptions (\cite{iqp_add2}).} (\cite{iqp_add2, iqp_add,iqp_mult}) for classical algorithms. For this reason, although the training can be offloaded to classical hardware, a quantum computer is needed to deploy and sample from the trained circuit at inference time, where real computational advantages may be present. 

The resulting classical algorithm implementing the MMD loss can be written in an efficient form that exploits basic linear algebra subroutines, which we implement in JAX (\cite{jax}) and train via automatic differentiation using the recently released software package \emph{IQPopt} (\cite{iqpopt}). This results in an algorithm that features linear scaling with respect to both the number of qubits and the number of gates, which can scale to circuits with thousands of qubits and millions of parameters using a single compute node of a computing cluster. The reliance on linear algebra subroutines also means that, like neural networks, the approach is a winner of the `hardware lottery' (\cite{hooker2021hardware}). The algorithm can therefore benefit from hardware accelerators such as graphics processing units or tensor processing units, and can in principle leverage the same multi-node distributed training methods that have been developed in the context of training large neural network models.

Although our models are similar to neural networks in terms of training hardware requirements, they differ from common classical generative models in a number of important ways. First, due to the use of qubit circuits, our generative models most naturally parameterise distributions over bitstrings rather than continuous vectors or large discrete alphabets. Second, since our models can be understood as implicit generative models\footnote{We note that even though inverse polynomial additive error estimates of probabilities are possible for our model class (\cite{pashayan2020estimation}), this is not precise enough to estimate log probabilities to a suitable precision which effectively renders the model implicit.} (\cite{mohamed2016learning}), we must train them using a suitable loss function, which we choose to be the MMD loss. This loss function has had some success in the classical literature (\cite{li2015generative,li2017mmd,binkowski2018demystifying}), but is not a common choice for modern models such as diffusion models or transformers, which train on loss functions more closely related to the log-likelihood. Third, it is still unclear what the limits of expressivity of the model class are, and we will show in Sec.~\ref{sec:coherence} that--at least if no ancilla qubits are used--the model is not a universal approximator of probability distributions over bitstrings.

To understand the potential of the approach we therefore conducted a number of numerical studies in which we trained circuits with up to one thousand qubits (Sec.~\ref{sec:experiments}). In order to evaluate the performance of the trained models we make use of two methods: the MMD with respect to a test set (\cite{o2021evaluation, lueckmann2021benchmarking, sutherland2016generative}) and the kernel generalised empirical likelihood (\cite{ravuri2023understandingdeepgenerativemodels}), both of which we can estimate at scale on classical hardware using the same mathematical tools used for training. As a comparison we also train two energy-based classical models: a restricted Boltzmann machine, and an energy based model whose energy function is a feedforward neural network. Encouragingly, the results show that it is possible to successfully train large-scale quantum models and obtain results that compete with or outperform the classical models when a similar amount of hyperparameter optimisation is performed on each model. 
For larger problems, the superior results of the quantum models are largely due to issues the classical models encountered during training, such as mode collapse or model imbalance. While much better results are likely possible for the classical models with further hyperparameter optimisation or tailored initialization strategies, this finding does suggest that parameterized IQP circuits can be trained with relative ease. We also compared the quantum models to trained classical generative models that are published in the genomics literature (\citet{yelmen2021creating}), which provides an additional independent comparison. Training on the same data, our quantum model performs similarly to the published classical models when evaluating the MMD with respect a test set. Although this is not an entirely fair comparison since we also train on the MMD, visual observation of the two-body correlations of the quantum model show that it has successfully learned a lot of the structure present in the data.  

Despite training large circuits, we were not hindered by problems of barren plateaus (\cite{mcclean2018barren,ragone2024lie,arrasmith2022equivalence}). This may have been because we employed a data-dependent parameter initialisation strategy in which parameters of two-qubit gates are initialised proportional to the corresponding covariance of the training data. Another possibility is that the model does not suffer from barren plateaus for the regime in which we trained, however this is still unclear (see Sec.~\ref{sec:bps} for more discussion on this point). We also show how the quantum model can be adapted in two ways. First, in Sec.~\ref{sec:bitflip} we show that it is possible to remove coherence from the quantum model, which results in an analogous classical model that we can also train at scale with a similar method. Despite undergoing identical hyperparameter optimisation and training strategies, we see that the decohered model fails to train on all but the smallest datasets, suggesting that coherence plays a critical role in the performance of the quantum model. Second, in Sec.~\ref{sec:sym} we show how to encode a specific global bitflip related symmetry into the model which matches the bias of one of the datasets we train on. The approach we use can most likely be generalized to construct quantum generative models whose distributions are invariant with respect to bitflip-type symmetries corresponding to powers of the group $\mathbb{Z}_2$. 

We believe our work offers a fresh perspective on the potential of variational approaches to quantum machine learning, and hopefully injects some much needed optimism regarding the prospect of scaling such models to large circuits. Aside from providing a scalable method for training, our work also opens up the possibility of heuristic studies into the behavior of quantum machine learning models at large scales, which may uncover insights beyond what is possible with purely theoretical analysis.


\section{Related work}\label{sec:related work}
The idea to use the maximum mean discrepancy to train (classical) generative models was first proposed in \cite{li2015generative} (see also \cite{li2017mmd,binkowski2018demystifying},), and was subsequently adapted to the quantum setting by \cite{liu2018differentiable}.  A recent in-depth study of the performance of the MMD for quantum generative models, as well as a discussion on the challenges of training quantum generative models can be found in \cite{rudolph2024trainability}. A general overview of learning in implicit generative models (a class to which our models belong)  can be found in \cite{mohamed2016learning}. 

A number of works have investigated using tools from classical simulability of quantum circuits to reduce the cost of training quantum generative models. \cite{kasture2023protocols} proposed a similar idea to ours, but relied on classical simulations of probabilities rather than expectation values, which limited them to training circuits of at most 30 qubits. \cite{rudolph2023synergistic} propose to use classical tensor network simulations in the initial stage of training, however this method still requires training on quantum hardware in order to reach models that cannot be simulated classically. \cite{fermionicml} propose to parameterise a class of matchgate circuits to construct scalable models (for classification) in a similar spirit to us, however the models they train are strongly simulable so cannot lead to a quantum advantage. Finally, the work of \cite{bako2024problem} shares some similarities with ours since they train a similarly structured circuit with the MMD, and encode graph based correlation structures as we also do in Sec.\ \ref{sec:exp:sf}, however training must be done on quantum hardware.

Regarding other approaches to scalability, a modern approach is to use the quantum computer for an initial data acquisition phase only, for example, in order to estimate a classical shadow (\cite{basheer2023alternating}) or a decomposition of the initial state in terms of the circuit's dynamical Lie algebra (\cite{goh2023lie}). Like ours, these approaches improve scalability by offloading the burden of training to a classical computer, however it is unclear how useful these methods can be. Classical shadows (\cite{huang2020predicting}) are limited to low body information, which may limit their use for high dimensional problems, and circuits with a suitably small dynamical Lie algebra are rare. Other approaches, like quantum kernel methods (\cite{schuld2021supervised}) or quantum principal component analysis (\cite{lloyd2014quantum}), also avoid variational optimisation on the quantum computer but encounter issues with scalability (with respect to the dataset size) and dequantisation algorithms (\cite{tang2021quantum}) respectively. 

\section{Generative learning with parameterised IQP circuits}\label{sec:prelims}
In this section we outline the setting of generative learning, introduce the class of quantum circuits that comprise our generative models, and describe the loss function we use to train the models.

\subsection{Generative learning}
The framework we consider in this work is that of \emph{generative learning}. We assume access to a dataset $\mathcal{X} = \{\x_i\}$ of vectors $\x_i$, called the \emph{training data}, where the $\x_i$ are sampled i.i.d.\ from a ground truth distribution $p(\x)$. Since we will be working with qubit-based quantum computers, we further assume the vectors $\x_i$ are bitstrings; $\x_i\in\{0,1\}^n$. A \emph{generative model} is a parameterised conditional distribution $q_{\btheta}(\x) \equiv q(\x \vert \btheta)$, where $\btheta$ is a vector of parameters to be inferred from the training data. The general aim of generative learning is to find a choice of parameters so that samples drawn from $q_{\btheta}$ closely resemble those of $p$. This is an intentionally vague definition, since the precise notion of `closeness' is not unique and may vary depending on the specific task the generative model is intended to solve. We will return to this issue when discussing model evaluation in Sec.\ \ref{sec:eval}. 

\subsection{Parameterised IQP circuits}
We will work with a class of quantum generative models that we call \emph{parameterised IQP circuits} due to their close connection with instantaneous quantum polynomial-time circuits (\cite{iqp_add,nakata2014diagonal}). 
\begin{definition}[parameterised IQP circuit]\label{def:iqpcircuits}
    A parameterised IQP circuit on $n$ qubits is a circuit comprised of the following:
    \begin{enumerate}
        \renewcommand{\labelenumi}{(\roman{enumi})}
        \item State initialisation in $\ket{0}$
        \item Parameterised gates of the form $\exp(i\theta_j X_{\g_j})$, where $X_{\g_j}$ is a tensor product of Pauli X operators acting on a subset of qubits specified by the nonzero entries of $\g_j\in\{0,1\}^n$
        \item Measurement in the computational basis
    \end{enumerate}
\end{definition}

The full parameterised unitary is thus $U(\btheta)=\prod_j \exp(i\theta_j X_{\g_j})$ where $\btheta = (\theta_1, \cdots, \theta_m)$ denotes the vector of trainable parameters. These circuits are particularly interesting from the standpoint of generative learning, since there exist examples of such circuits for which there is no efficient classical algorithm to sample from the output distribution up to either additive (\cite{iqp_add,iqp_add2}) or multiplicative (\cite{iqp_mult}) error, assuming plausible conjectures in complexity theory hold. It therefore seems unlikely that it is possible to construct classical generative models that closely emulate the behavior of these circuits, which opens the possibility of uniquely quantum advantages. 

\subsection{The maximum mean discrepancy}
A common approach to training generative models is to define a differentiable loss function that quantifies the performance of the model on training data, and infer parameters via a gradient-descent based method. This is the approach we will adopt to train our quantum models. However, unlike many classical generative models, our models are implicit generative models (\cite{mohamed2016learning}), meaning that we cannot estimate probabilities to high precision, and we therefore need to use a loss function that is compatible with such models. 

The loss function we will use is is based on the \emph{maximum mean discrepancy} (MMD) (\cite{JMLR:v13:gretton12a}). This loss has had some success in the classical machine learning literature (\cite{li2015generative, sutherland2016generative}), is suitable for implicit generative models, and is one of the few scalable methods known for training quantum generative models (\cite{rudolph2024trainability}). The MMD is a type of distance function between distributions (called an integral probability metric), and corresponds to the 2-norm between the expected feature vectors under a feature map $\phi$ defined via a kernel function $k(\x,\y)=\langle \phi(\x)\vert \phi(\y)\rangle$:
\begin{align}
    \MMD(p, q_{\btheta}) = \vert\vert \mathbb{E}_{\x\sim p}[\phi(\x)] - \mathbb{E}_{\y\sim q_{\btheta}}[\phi(\y)]\vert\vert,
\end{align}
where $\vert\vert \x \vert\vert = \sqrt{\x^* \cdot\x}$. To obtain our loss function, we take the square of the \MMD, and substitute the kernel function from the resulting inner products.
\begin{definition}[MMD loss function]\label{def:mmd}
The squared maximum mean discrepancy between distributions $p$ and $q_{\btheta}$ is
    \begin{align}\label{eq:truemmd}
   {\normalfont{\text{MMD}}}^2(p, q_{\btheta}) = \mathbb{E}_{\x,\y\sim p}\left[k(\x,\y)\right] - 2\mathbb{E}_{\x\sim p, \y\sim q_{\btheta}}\left[k(\x,\y)\right] +  \mathbb{E}_{\x,\y\sim q_{\btheta}}\left[k(\x,\y)\right].
\end{align}
\end{definition}
We will work with a common choice of kernel function, namely the \emph{Gaussian kernel} (or radial basis function kernel):
\begin{align}\label{eq:kernel}
    k(\x, \y) = \exp\left(\frac{-\vert\vert \x-\y \vert\vert^2}{2\sigma^2}\right).
\end{align}
Since this kernel is characteristic, it follows that one has MMD$^2=0$ iff $p=q_{\btheta}$ (\cite{JMLR:v13:gretton12a}). The parameter $\sigma$, called the \emph{bandwidth}, should be chosen carefully from data to provide a meaningful comparison\footnote{We remark that we choose $\sigma$ to correspond to the standard deviation of the (unnormalized) Gaussian distribution given by \eqref{eq:kernel}, however in some works (such as \cite{rudolph2024trainability}) $\sigma$ corresponds instead to the variance.} (\cite{sutherland2016generative}). A popular choice that we will make use of is the \emph{median heuristic} (\cite{medianheuristic}).
\begin{definition}[median heuristic]
    Given a dataset $\{\x_i\}$, the median heuristic $\hat{\sigma}$ is the median value of pairwise distances of the points:
\end{definition}
\begin{align}
    \hat{\sigma} = \text{med}( \{ \vert\vert \x_i -\x_j \vert\vert\}_{i,j}).
\end{align}
The most widely used way to compute unbiased estimates of the MMD$^2$ is to sample batches of vectors $\mathcal{X}=\{\x_i\sim p\}$ and $\mathcal{Y}=\{\y_j\sim q_{\btheta}\}$ and use the estimator (see \cite{JMLR:v13:gretton12a})
\begin{multline}\label{eq:MMD_greton}
    \hat{\text{MMD}}^2(\mathcal{X},\mathcal{Y}) = \frac{1}{|\mathcal{X}|(|\mathcal{X}|-1)} \sum_{i\neq j} k(\x_i, \x_j) - \frac{2}{|\mathcal{X}||\mathcal{Y}|}\sum_{i,j}k(\x_i,\y_j)\\ + \frac{1}{|\mathcal{Y}|(|\mathcal{Y}|-1)}\sum_{i\neq j}k(\y_i,\y_j).
\end{multline}
This estimator can be shown to be unbiased so that its expected value coincides with the true MMD$^{2}$, i.e.\ we have
\begin{align}
    \mathbb{E}_{\{\x_i\sim p\},\{\y_j\sim q_{\btheta}\}}\left[\hat{\text{MMD}}^2(\{\x_i\},\{\y_j\})\right] = \text{MMD}^2(p, q_{\btheta}).
\end{align}
Unbiased estimates of the gradient of the MMD$^2$ for parameterised quantum circuits can be found by differentiating the expression \eqref{eq:truemmd} and using the parameter shift rule (see \cite{liu2018differentiable} App.~A), which can be used to train the model. Estimating the gradient in this way however requires sampling from at least two circuits for every trainable parameter (\cite{wierichs2022general, kyriienko2021generalized, vidal2018calculus}), making scaling to very large circuits prohibitively expensive in practice. 

\section{Efficient training on classical hardware}\label{sec:training}
In this section we will see that, when the quantum generative model is a parameterised IQP circuit, it is possible to construct unbiased estimators of the MMD$^2$ and its gradient using an efficient classical algorithm. This will follow from the combination of two ingredients: (i) a proof by \citet{nest2010simulating} showing how to classically estimate expectation values of Pauli Z words for IQP circuits, and (ii) a recent observation by \citet{rudolph2024trainability} showing that the MMD$^2$ can be expressed as a probabilistic mixture of these expectation values. By using the classical estimation algorithm within this form of the MMD$^2$, we arrive at a classical algorithm to estimate and optimize the loss.  

\subsection{Efficient estimation of expectation values}
In this section we concentrate on the problem of estimating expectation values of Pauli Z words applied at the output of parameterised IQP circuits. That is, we wish to estimate quantities.
\begin{align}\label{eq:zexpvals}
    \langle Z_{\a} \rangle_{q_{\btheta}} = \bra{0} U^{\dagger}(\btheta) Z_{\a} U(\btheta)\ket{0},
\end{align}
where $\ket{\psi(\btheta)}=U(\btheta)\ket{0}$, and $Z_{\a}$ is a tensor product of Pauli Z operators specified by the non-zero positions of $\a\in\{0,1\}^n$. That is,
\begin{align}
    Z_{\a} = \prod_{i=1}^n (Z_i)^{a_i},
\end{align}
where $Z_i$ denotes a Pauli Z operator acting on qubit $i$. A classical algorithm for this task is implied by the results of \cite{nest2010simulating} (Thm.~3) but we repeat it here in a simplified form that applies directly to our circuits. 

\begin{proposition}[\cite{nest2010simulating}]\label{prop:nest}
    Given a parameterised IQP circuit $q_{\btheta}$, an expectation value $\langle Z_{\a} \rangle_{q_{\btheta}}$ and an error $\epsilon=\text{poly}(n^{-1})$, there exists a classical algorithm that requires poly$(n)$ time and space, and samples a random variable with standard deviation less than $\epsilon$ that is an unbiased estimator of $\langle Z_{\a} \rangle_{q_{\btheta}}$. 
\end{proposition}

\noindent\textit{Proof}: Inserting identities $\mathbb{I}=H^2$, where $H$ is an $n$-fold tensor product of Hadamard unitaries, between operators in \eqref{eq:zexpvals} the expression becomes
\begin{align}
    &\bra{0}H (H U^{\dagger}(\btheta) H)(H Z_{\a} H)(H U(\btheta)H)H\ket{0} \nonumber \\ &\quad\quad= 
    \bra{+}^{\otimes n} D^{\dagger}(\btheta) X_{\a} D(\btheta)\ket{+}^{\otimes n}, \label{eq:bowles_hadamard}
\end{align}
with $D(\btheta)$ a diagonal unitary analogous to $U(\btheta)$ where the generators are now Pauli Z tensors,
\begin{align}\label{eq:diagcircuit}
    D(\btheta)=\prod_{j}e^{i\theta_j Z_{\g_j}}.
\end{align}
Since $D(\btheta)$ is diagonal, its eigenvectors are computational basis states. The corresponding eigenvalues $\lambda_{\z}$ are easy to compute since a basis state $\ket{\z}$ picks up a phase $\exp(i\theta_{j}(-1)^{\g_{j}\cdot\z})$ for each gate in \eqref{eq:diagcircuit}:
\begin{align}\label{eq:eigenvals}
    D(\btheta)\ket{\z}=\exp(i\sum_{j} \theta_{j}(-1)^{\g_{j}\cdot\z})\ket{\z} = \lambda_{\z}\ket{z}.
\end{align}
Writing $\ket{+}^{\otimes n}=2^{-n/2}\sum_{\z}\ket{\z}$ in \eqref{eq:bowles_hadamard} and using the above and $X_{\a}\ket{\z}=\ket{\z\oplus {\a}}$ we arrive at 
\begin{align}
     \langle Z_{\a} \rangle_{q_{\btheta}} = \frac{1}{2^n} \sum_{\z} \lambda^{*}_{\z\oplus{\a}}\lambda_{\z} 
    =\frac{1}{2^n} \sum_{\z}\text{Re}\left[\lambda^{*}_{\z\oplus\a}\lambda_{\z}\right], \label{eq:zaq}
\end{align}
where we have have taken the real component since $\langle Z_{\a} \rangle_{q_{\btheta}}$ is guaranteed to be real. Substituting the expression for $\lambda_{\z}$ from \eqref{eq:eigenvals} and using the fact that \eqref{eq:zaq} is an expectation value with respect to the uniform distribution $U$ over bitstrings we then find
\begin{align}\label{eq:expvalexp}
    \langle Z_{\a} \rangle_{q_{\btheta}} =\mathbb{E}_{\z\sim U}\Big[ \cos\Big(\sum_j \theta_{j}(-1)^{\g_{j}\cdot \z}(1-(-1)^{\g_j\cdot \a}\Big) \Big].
\end{align}
The above now allows us to compute unbiased estimates of $\langle Z_A \rangle_{q_{\btheta}}$ efficiently by replacing the expectation with an empirical mean. That is, if we sample a batch of bitstrings $\mathcal{Z}=\{\z_i\}$ from the uniform distribution and compute the sample mean
\begin{align}\label{eq:expvalsample}
   \hat{\langle Z_{\a}\rangle}_{q_{\btheta}} = \frac{1}{|\mathcal{Z}|}\sum_{i=1}^{|\mathcal{Z}|}\cos\Big(\sum_j \theta_j(-1)^{\g_j\cdot \z_i}(1-(-1)^{\g_j\cdot \a})\Big),
\end{align}
we obtain an estimate of $\langle Z_{\a}\rangle_{q_{\btheta}}$. One sees this estimate is unbiased since 
\begin{align}
    \mathbb{E}_{\mathcal{Z}}[\hat{\langle Z_{\a}\rangle}_{q_{\btheta}}] = \langle Z_{\a}\rangle_{q_{\btheta}}
\end{align}
by virtue of \eqref{eq:expvalsample} being a mean with respect to the elements of $\mathcal{Z}$. Since ${\langle Z_{\a}\rangle}_{q_{\btheta}}$ is an expectation with respect to a random variable bouned in $[-1,1]$, it follows that the variance of the sample mean \eqref{eq:expvalsample} is bounded by $1/|\mathcal{Z}|$. Taking $|\mathcal{Z}|\geq 1/\epsilon^2=\text{poly}(n)$ we therefore obtain an estimator for $\langle Z_{\a} \rangle_{q_{\btheta}}$ with standard deviation less than $\epsilon$. $ \blacksquare$

\subsection{\MMDD\, as a mixture of expectation values}
The second ingredient we need comes from writing the MMD$^2$ in terms of expectation values rather than probabilities. In particular, in the work of \cite{rudolph2024trainability} (see App.~C) they show the following.
\begin{proposition}[\cite{rudolph2024trainability}]\label{prop:mmd}
The squared maximum mean discrepancy between distributions $p$ and $q_{\btheta}$ can be expressed as 
    \begin{align}\label{eq:expmmd}
    {\normalfont{\text{MMD}}}^2(p,q_{\btheta}) 
    &= \mathbb{E}_{\a\sim \mathcal{P}_{\sigma}(\a)}\Big[ \big(\langle Z_{\a} \rangle_p  - \langle Z_{\a} \rangle_{q_{\btheta}}\big)^2 \Big]
\end{align}
where $\a$ is distributed according to a product of Bernoulli distributions with probability $p_{\sigma}$,
\begin{align}
    \mathcal{P}_{\sigma}(\a) = (1-p_{\sigma})^{n-\vert \a \vert}p_{\sigma}^{\vert \a \vert}; \quad\quad p_{\sigma}=\frac{1-e^{-\frac{1}{2\sigma^2}}}{2},
\end{align}
and where $\vert \a \vert$ is the Hamming weight of $\a$. 
\end{proposition}
 In the above we have that 
\begin{align}
   \langle Z_{\a} \rangle_p = \mathbb{E}_{\x\sim p}\left[ (-1)^{\x\cdot \a }\right]
\end{align}
is the usual expectation value of the corresponding $\pm 1$ valued variables on the subset of bits where $a_i\neq 0$. Note that we can construct an unbiased estimate $\langle \hat{Z}_{\a} \rangle_p$ of $\langle Z_{\a} \rangle_p$ by sampling a batch of data $\mathcal{X}=\{\x_i\sim p\}$ and computing the sample mean:
\begin{align}\label{eq:zax}
   \langle \hat{Z}_{\a} \rangle_p = \frac{1}{|\mathcal{X}|}\sum_i (-1)^{\x_i\cdot \a}.
\end{align}
There are two important things to notice from Prop.~\ref{prop:mmd}. First, when minimising the MMD$^2$ one is effectively attempting to match the expectation values of the two distributions, where the importance of given expectation value depends on the probability $\mathcal{P}_{\a}$ of it being sampled. Second, the distribution over $\a$ is dependent on the bandwidth parameter $\sigma$ of the kernel, and is such that the value $\vert \a \vert$ will be peaked around the mean value $n p_\sigma$ by virtue of it being distributed binomially. One sees that lower values of $\sigma$ result in larger average values of $\vert \a \vert$, so that lower bandwidths effectively probe higher order correlations. This can have practical implications for training, as we discuss in Sec.~\ref{sec:trainstrat}. 

\subsection{Unbiased estimates of the \MMDD}
With Prop.~\ref{prop:nest} and \ref{prop:mmd} in hand we are now ready to construct estimates of the MMD$^2$ for parameterised IQP circuits. 
A straightforward strategy is the following.
\begin{enumerate}
    \item Sample batches of bitstrings 
    \begin{align}
        \mathcal{X} = \{\x_i \sim p\},\; \mathcal{A} = \{\a_j\sim\mathcal{P}_\sigma\},\; \mathcal{Z} = \{\z_k \sim U\} . \nonumber
    \end{align}
    \item For each $\a_j \in \mathcal{A}$ use \eqref{eq:expvalsample} and \eqref{eq:zax} to obtain estimates $\langle \hat{Z}_{\a_j} \rangle_{q_{\btheta}}$, $\langle \hat{Z}_{\a_j} \rangle_{p}$ of $\langle Z_{\a_j} \rangle_{q_{\btheta}}$ and $\langle Z_{\a_j} \rangle_{p}$.
    \item Estimate the MMD$^2$ by converting \eqref{eq:expmmd} to a sample mean, i.e.
    \begin{align}
    \hat{\text{MMD}^2}(\mathcal{X, A, Z}, \btheta) =  \frac{1}{\vert \mathcal{A} \vert}\sum_j\big(\hat{\langle Z_{\a_j} \rangle}_p  - \hat{\langle Z_{\a_j} \rangle}_{q_{\btheta}}\big)^2 .
    \end{align}
\end{enumerate}
This method does not give an unbiased estimator however due to the quadratic nonlinearity. In App.~\ref{app:unbiased} we show how to modify the expression to arrive at an unbiased estimator $\hat{\text{MMD}_{\text{u}}^2}$ that we use to train our models, which we present in the following proposition. 

\begin{proposition}[Unbiased estimates of the MMD$^2$]
Given a dataset $\mathcal{X}=\{\x_i \sim p\}$ and batches of bitstrings $\mathcal{A}=\{\a_i\sim\mathcal{P}_\sigma\}, \mathcal{Z} = \{\z_i \sim U\}$, the estimator 
\begin{align}\label{eq:mmd_beast}\hat{{\normalfont{\text{MMD}}}_{\text{u}}^2}(\mathcal{X},\mathcal{A}, \mathcal{Z}, \btheta) & = \frac{1}{|\mathcal{A}| |\mathcal{Z}|(|\mathcal{Z}|-1)}\sum_{i,j,k\neq j}f(\a_i,\z_j, \btheta)f(\a_i,\z_k, \btheta) \nonumber \\ &\quad\quad - \frac{2}{|\mathcal{A}| |\mathcal{Z}|  |\mathcal{X}|}\sum_{i,j,k}f(\a_i,\z_j, \btheta)(-1)^{\x_k\cdot \a_i} \nonumber \\ &\quad\quad\quad\quad + \frac{1}{|\mathcal{A}| |\mathcal{X}|( |\mathcal{X}|-1)}\sum_{i,j,k\neq j}(-1)^{\x_j\cdot \a_i}(-1)^{\x_k\cdot \a_i},
\end{align}
where
\begin{align}
    f(\a,\z,\btheta) = \cos \left(\sum_j \theta_j (-1)^{\g_j \cdot \z}(1-(-1)^{\g_j\cdot \a})\right),
\end{align} 
is an unbiased estimator of ${\normalfont{\text{MMD}}}^2(p,q_{\btheta})$, where $q_{\btheta}$ is the distribution generated by a parameterised IQP circuit with parameters $\btheta$ and gates $\{\g_j\}$. That is, we have 
\begin{align}
    \mathbb{E}_{\mathcal{X, A, Z}}\left[ \hat{{\normalfont{\text{MMD}}}_{\text{u}}^2}(\mathcal{X, A, Z}, \btheta)\right] = {\normalfont{\text{MMD}}}^2(p,q_{\btheta}).
\end{align}
\end{proposition}

\subsection{Training via automatic differentiation}
Note that the expression \eqref{eq:mmd_beast} is a differentiable function of $\btheta$ and a deterministic function of $\mathcal{A}, \mathcal{Z}, \mathcal{X}$ (for fixed $\btheta$). As a result, we can obtain an estimate of the gradient $\nabla_{\btheta}\text{MMD}^2(p,q_{\btheta})$ by first sampling batches $\mathcal{A}, \mathcal{Z}, \mathcal{X}$, computing an estimate of $\hat{\text{MMD}_{\text{u}}^2}$ via \eqref{eq:mmd_beast}, and then computing its gradient via automatic differentiation\footnote{We note one could also differentiate the expression by hand, but we chose to use automatic differentiation to maintain code flexibility.} (\cite{griewank2008evaluating}). This results in an unbiased estimate of the gradient since by averaging over the sampling of batches $\mathcal{A}, \mathcal{Z}, \mathcal{X}$,
\begin{align}
    \mathbb{E}_{\mathcal{A}, \mathcal{Z}, \mathcal{X}}[\nabla_{\btheta}\hat{\text{MMD}_{\text{u}}^2}(\mathcal{A}, \mathcal{Z}, \mathcal{X}, \btheta)] =  \nabla_{\btheta}\mathbb{E}_{\mathcal{A}, \mathcal{Z}, \mathcal{X}}[\hat{\text{MMD}_{\text{u}}^2}(\mathcal{A}, \mathcal{Z}, \mathcal{X}, \btheta)] = \nabla_{\btheta}\text{MMD}^2(p,q_{\btheta}).
\end{align}
The expression \eqref{eq:mmd_beast} is also a relatively simple mathematical expression that employs basic linear algebra. Because of this, one can exploit fast subroutines for matrix multiplication to evaluate estimates of the MMD$^2$ and its gradient in a way that is highly efficient, and can accelerated with access to graphics processing units. This is implemented in the python package \emph{IQPopt} (\cite{iqpopt}), which was developed alongside this project in order train our models. The package is able to compute estimates of the MMD$^2$ and its gradient for parameterised IQP circuits in JAX (\cite{jax}), and trains the parameters via gradient descent methods implemented via JAXopt (\cite{jaxopt}). For more information about the IQPopt package and the computational techniques employed, we refer the reader to \citet{iqpopt}.

\section{Stochastic bitflip circuits and the role of coherence}\label{sec:coherence}
In this section we investigate the role of coherence in parameterised IQP circuits. As we will see, this will lead us to a classical model that can be seen as a decohered version of the quantum circuit, which we can also train with a similar method. We then use these tools to discuss expressivity and universality of the model class. 

\subsection{Expectation values of parameterised IQP circuits}

We first derive an expression for expectation values $\langle Z_{\a}\rangle$ of parameterised IQP circuits. Note that since $Z \exp(i\theta X) = \exp(-i\theta X) Z$ due to anticommutation of $Z,X$ we have
\begin{align}
    \langle Z_{\a}\rangle = \bra{0}U^\dagger(\btheta)Z_{\a}U(\btheta)\ket{0} &=\bra{0}\prod_{j\in\mathcal{S}_{\a}}\exp(-2i\theta_j X_{\g_j})\ket{0} \\
    &=\bra{0}\prod_{j\in\mathcal{S}_{\a}}(\cos(2\theta_j)\mathbb{I}+i\sin(-2\theta_j)X_{\g_j})\ket{0},
\end{align}
where $\mathcal{S}_{\a}=\{j\vert \{X_{\g_j},Z_{\a}\}\}=0$ specifies the generators that anticommute with $Z_{\a}$. From this we can see that the only terms that survive when expanding the product are those that result in an identity operator. We thus have 
\begin{align}\label{eq:expval_intefere}
    \langle Z_{\a} \rangle = \prod_{j\in\mathcal{S}_{\a}} \cos(2\theta_j) + \sum_{\omega \in \Omega}\prod_{\substack{j\in \mathcal{S}_{\a} \\ j \notin \omega}}\cos(2\theta_j)\prod_{j \in \omega}i\sin(-2\theta_j)
\end{align}
where $\Omega = \{\omega_1, \omega_2,\cdots\}$ is the collection of sets of indices $\omega\subseteq \mathcal{S}_{\a}$ such that $\prod_{j\in \omega} X_{\g_j} = \mathbb{I}$ holds for each $\omega\in\Omega$. In general, there will be an exponential number of elements (in $n$) in $\Omega$, which prevents an efficient brute force calculation. 

\subsection{Removing coherence: stochastic bitflip circuits}\label{sec:bitflip}
We now describe a classical model that is equivalent to a parameterised IQP circuit in which one uses decohered classical versions of the parameterised gates. We can also derive a similar expression for the expectation values of these circuits, which will help us to clearly delineate the role of coherence in the quantum models. 

To understand the classical model, we first note that the action of the quantum gate $\exp(i \theta_j X_{\g_j})$ on a computational state $\ket{\x}$ is
\begin{align}
    \exp(i \theta_{_j} X_{\g_j})\ket{\x} = \cos(\theta_j)\ket{\x}+i\sin(\theta_j)\ket{\x\oplus\g_j}.
\end{align}
That is, the gate flips some of the bits of $\x$ in coherent superposition such that the probability of observing the flipped bitstring $\x\oplus\g_j$ from a subsequent measurement is $\sin^2(\theta_j)$. We construct our classical model by considering an incoherent version of this gate, which flips the same bits in a classical stochastic manner:
\begin{align}\label{eq:bitflip}
    \ket{\x}\bra{\x} \rightarrow \cos^2(\theta_j)\ket{\x}\bra{\x} + \sin^2(\theta_j)\ket{\x\oplus\g_j}\bra{\x\oplus\g_j}.
\end{align}
We call circuits constructed from these gates \emph{stochastic bitflip circuits}, and in analogy to Def.~\ref{def:iqpcircuits} define them as follows. 
\begin{definition}[Stochastic bitflip circuit]\label{def:bitflip}
    A stochastic bitflip circuit on $n$ bits is a classical stochastic circuit comprised of the following:
    \begin{enumerate}
        \renewcommand{\labelenumi}{(\roman{enumi})}
        \item Initialisation of the all zero bitsting $(0,\cdots,0)$
        \item Stochastic parameterised gates of the form \eqref{eq:bitflip} that flip subsets of bits
        \item Read out of the final bitstring
    \end{enumerate}
\end{definition}

We can derive an expression for expectation values $\langle Z_{\a} \rangle$ for these circuits by noting that they can be constructed as special cases of parameterised IQP circuits. In particular, given a stochastic bitflip circuit on $n$ bits with $m$ gates $\{\g_j\}$, construct a parameterised IQP circuit on $n+m$ qubits that has gate generators $X_{\g_j} X_{n+j}$ (that is, each generator has an $X$ operator on a unique ancilla qubit). Consider expectation values $\langle Z_{\a}\rangle$ of this parameterised IQP circuit, where $Z_{\a}$ acts nontrivially on the first $n$ qubits only. The action of a gate with generator $X_{\g_j} X_{n+j}$ on state $\ket{\x}\ket{0}^{j}$ is
\begin{align}
    \cos(\theta_j)\ket{\x}\ket{0}^{(j)} + i\sin(\theta_j)X_{\g_j}\ket{\x}\ket{1}^{(j)}.
\end{align}
Since we are interested in expectation values of the first $n$ qubits only, we may  trace out the ancilla qubit, so that the action on $\ket{\x}$ is
\begin{align}
    \cos^2(\theta_j)\ket{\x}\bra{\x} + \sin^2(\theta_j)X_{\g_j}\ket{\x}\bra{\x}X_{\g_j} =  \cos^2(\theta_j)\ket{\x}\bra{\x} + \sin^2(\theta_j)\ket{\x\oplus\g_j}\bra{\x\oplus\g_j},
\end{align}
i.e. a stochastic bit flip of the form \eqref{eq:bitflip}. Expectation values evaluated on the first $n$ qubits of this circuit are therefore the same as the corresponding stochastic bitflip circuit, and we may thus use \eqref{eq:expval_intefere}. Noting that each of the generators of the constructed parameterised IQP circuit contains an $X$ operator on a unique ancilla qubit, we have that $\Omega$ must be the empty set and so
\begin{align}\label{eq:expval_bitflip}
    \langle Z_{\a} \rangle = \prod_{j\vert \{X_{\g_j},Z_{\a}\}=0} \cos(2\theta_j)
\end{align}
for stochastic bitflip circuits. This makes for a clear exhibition of the role of coherence: in \eqref{eq:expval_intefere} the first term can be understood as a classical term that corresponds to the equivalent stochastic bitflip circuit, whereas the remaining terms are the result of coherence. The form of \eqref{eq:expval_bitflip} also means we can estimate the MMD$^2$ and its gradient efficiently in a similar manner to parameterised IQP circuits, which we also implement in the IQPopt package (\cite{iqpopt}). This allows us to train both the parameterised quantum model and its classical bitflip surrogate model, which can help understand if coherence plays a role in model performance. In the experiments of Sec.~\ref{sec:experiments} we will take this approach and see that indeed coherence appears to dramatically affect model performance. In App.~\ref{app:toy} we also use the above to present a toy example where coherence is provably beneficial for a certain type of expressivity.  

\subsection{Limits of expressivity}\label{sec:expressivity}
An important feature to understand in any class of generative models is expressivity, which relates to the set of possible distributions that can be prepared (or approximately prepared) by instances of models from the class (\cite{schuld2021effect, gil2024expressivity, wu2021expressivity, shin2023exponential}). In particular, many classical generative model classes are known to exhibit universality, in the sense that any valid probability distribution can be prepared by an appropriate model in the class. Universality is generally a desirable property for a model class to have, since it implies that the class is flexible enough to learn any distribution, although this does not mean that learning will be efficient. 

One may wonder whether the class of parameterised IQP circuits is universal for probability distributions over bitstrings. That is, for any valid distribution $p(\vec{x})$ over bitstrings $\x$, does there exists a parameterized IQP circuit and choice of parameters such that $\vert \bra{x}U(\btheta)\ket{0}\vert^2 = p(\vec{x})$. One may suspect this to be the case since the maximum number of free parameters in a parameterised IQP circuit of $n$ qubits is $2^n-1$ (corresponding to all possible gate choices), which is precisely the same number of free parameters needed to describe a general distribution $p(\vec{x})$ of $n$ bits. Despite this, it turns out that the model class is not universal in this sense. More specifically, parameterised IQP circuits on $n$ qubits cannot capture the full space of distributions on $n$ bits. This can in fact be seen already for the case $n=2$ (see App.~\ref{app:universality}) by showing that two-qubit parameterised IQP circuits are equivalent to stochastic bitflip circuits, which are easily seen to be non-universal. It is still possible that universality could be achieved via the use of ancillas, that is, by considering marginal distributions of circuits with greater than $n$ qubits, and understanding if this is the case would help evaluate the potential of parameterized IQP circuits as generative learning models. 

\section{Beyond IQP: incorporating symmetry into the ansatz}\label{sec:sym}
Constructing a machine learning model with good generalisation capability often relies of the ability to incorporate an inductive bias into the model that mirrors known structures that are present in the data (\cite{bowles2023contextuality, adam2019no, bronstein2021geometric}). The paradigmatic example of this is the convolutional neural network, whose widespread success at computer vision tasks can be attributed to the fact that the convolutional layers are built upon a translation symmetry (translation equivalence) that is often reflected in image data. In this section we show how a particular symmetry can be built into circuits with IQP structures. This will be achieved by modifying the input state, so that technically speaking, the circuits no longer belong to the class of IQP circuits as defined by Def.~\ref{def:iqpcircuits}. 

The symmetry in question is a particular invariance to global bitflips, whose corresponding group is $\mathbb{Z}_2$. In particular, we will construct a generative model $q_{\btheta}(\x)$ that respects the probabilistic invariance 
\begin{align}\label{eq:bias}
    q_{\btheta}(\x) = q_{\btheta}(\bar{\x}) \quad \quad \forall\;\btheta, \x
\end{align}
where $\bar{\x}$ is obtained from $\x$ by flipping all of the bits. In order to achieve this, we change the initial state $\ket{0}$ to a state that is an eigenstate of $\tilde{X} = X\otimes\cdots\otimes X$, the operator that corresponds to flipping all bits. Such a state is the GHZ state
\begin{align}
    \ket{\phi} = \frac{1}{\sqrt{2}}\left(\ket{0^{\otimes n}} + \ket{1^{\otimes n}}\right).
\end{align}
With this choice we see that
\begin{multline}
    p(\x\vert\btheta) = \vert\bra{\x}U(\btheta)\ket{\phi}\vert^2 = \vert\bra{\x}U(\btheta)\tilde{X}\ket{\phi}\vert^2 = \vert\bra{\x}\tilde{X}U(\btheta)\ket{\phi}\vert^2 = \vert\bra{\bar{\x}}U(\btheta)\ket{\phi}\vert^2 \\ = p(\bar{\x}\vert\theta) \quad \forall \theta
\end{multline}
as desired. Although the state $\ket{\phi}$ cannot be prepared by a parameterised IQP circuit, by expanding the expression and inserting Hadamards as in \eqref{eq:bowles_hadamard} we have
\begin{align}
    \bra{\phi}U^{\dagger}(\btheta)Z_{\a}U(\btheta)\ket{\phi} 
    &= \frac{1}{2}(\bra{0}^{\otimes n}U^{\dagger}(\btheta) Z_{\a} U(\btheta)\ket{0}^{\otimes n} + \bra{1}^{\otimes n} U^{\dagger}(\btheta) Z_{\a} U(\btheta)\ket{1}^{\otimes n} \nonumber \\ & + 2\text{Re}[\bra{0}^{\otimes n}U^{\dagger}(\btheta) Z_{\a} U(\btheta)\ket{1}^{\otimes n}]) \nonumber \\
    &= \frac{1}{2}(\bra{+}^{\otimes n}D^{\dagger}(\btheta) X_{\a} D(\btheta)\ket{+}^{\otimes n} + \bra{-}^{\otimes n} D^{\dagger}(\btheta) X_{\a} D(\btheta)\ket{-}^{\otimes n} \nonumber \\ & + 2\text{Re}[\bra{+}^{\otimes n}D^{\dagger}(\btheta) X_{\a} D(\btheta)\ket{-}^{\otimes n}])  \label{eqn:ZA}
\end{align}
and we can therefore approximate the expectation value by approximating each of the three terms in \eqref{eqn:ZA}. To do this we first note that since $\ket{-}^{\otimes n} = \sum_{\z} (-1)^{|\z|}\ket{\z}$ (where $|\z|$ is the Hamming weight of $\z$) it follows that
\begin{align}
D(\btheta)\ket{-}^{\otimes n} = \prod_{j=1}^{m}e^{i\theta_j Z_{\g_j}}\ket{-}^{\otimes n}
    &= \frac{1}{2^{n/2}}\sum_{\z} (-1)^{|\z|} \prod_{j=1}^{m}e^{i\theta_j (-1)^{\g_j\cdot\z}}\ket{\z}\\
    &= \frac{1}{2^{n/2}}\sum_{\z} (-1)^{|\z|} \lambda_{\z}\ket{\z}.
\end{align}
Carrying this extra minus sign into \eqref{eqn:ZA} we find
\begin{align}
\bra{\phi}^{\otimes n}U^{\dagger}(\btheta)Z_{\a}U(\btheta)\ket{\phi}^{\otimes n}  &= \frac{1}{2^n} \sum_{\z} \left(\frac{1}{2}\lambda_{\z}\lambda^*_{\z\oplus\a} + \frac{1}{2}(-1)^{|\z|+|\z\oplus\a|}\lambda_{\z}\lambda^*_{\z\oplus\a}+\text{Re}[(-1)^{|\z|}\lambda_{\z}\lambda^*_{\z\oplus\a}]\right) \nonumber \\ 
&= \frac{1}{2^n} \sum_{\z} \left(\frac{1}{2} + \frac{1}{2}(-1)^{|\a|}+(-1)^{|\z|}\right)\text{Re}[\lambda_{\z}\lambda^*_{\z\oplus\a}], \label{eqn:spin_sym}
\end{align}
where we have taken the real part of the entire expression in the second line since the expectation value is guaranteed to be real. Converting $\text{Re}[\lambda_{\z}\lambda^*_{\z\oplus\a}]$ to a cosine we arrive at 
\begin{align}
    \bra{\phi}U^{\dagger}(\btheta)Z_{\a}U(\btheta)\ket{\phi} = \mathbb{E}_{\z\sim U}\Big[\left(\frac{1}{2} + \frac{1}{2}(-1)^{|\a|}+(-1)^{|\z|}\right) \cos\Big(\sum_j \theta_{j}(-1)^{\g_{j}\cdot \z}(1-(-1)^{\g_j\cdot \a}\Big) \Big]
\end{align}
for which we can construct an unbiased estimate via a sample mean in the same fashion as \eqref{eq:expvalsample}. In Sec.~\ref{sec:experiments} we will study a dataset that features the invariance \eqref{eq:bias} and train a symmetrised model with this method. We note that the above approach suggests a general method to construct ans\"{a}tze with other symmetries beyond the simple one studied here by the following recipe:
\begin{enumerate}
    \item Identify a symmetry operator analogous to $\tilde{X}$ that commutes with all IQP gates (i.e.\ that is diagonal in the $X$ basis)
    \item Construct an initial state that is an eigenstate of the symmetry operator and has an efficient expansion in the computational basis
    \item Expand the expression for expectation values as in \eqref{eqn:ZA} and simplify the expression as an expectation over bitstrings
\end{enumerate} 
Using this recipe we expect that symmetries corresponding to the groups $\mathbb{Z}_2^k$ (i.e.\ bit-flipping symmetries) can be constructed if suitable initial states can be found. 

\section{Evaluating model performance}\label{sec:eval}
Evaluating the performance of a trained generative model is in general a difficult task due to the difficulty of working in exponentially large spaces (\cite{theis2015note, betzalel2022study, stein2024exposing}). There exist a wide variety of metrics (\cite{bischoff2024practical, alaa2022faithful}), however good performance on one metric does not necessarily imply good performance on another (\cite{theis2015note}). In this section we cover three evaluation metrics that we will use to evaluate the performance of parameterized IQP circuits in the numerical experiments that follow. 

\subsection{Evaluation metrics}
Perhaps the least controversial and widely used metric is the log likelihood of a test set. 
\begin{definition}[log likelihood of a test set] 
    Given a set of test data $\mathcal{X}_{\text{test}}=\{\x_i\}$, the log likelihood with respect to a generative model $q_{\btheta}$ is 
    \begin{align}
        LL(\mathcal{X}_{\text{test}}\vert q_{\btheta}) = \log\Big(\prod_i q(\x_i\vert \btheta)\Big) = \sum_i \log\Big(q(\x_i\vert \btheta)\Big). 
    \end{align}
\end{definition}
This has the simple interpretation as the log probability that the generative model produce the test data, and larger values therefore imply better performance from a maximum likelihood perspective. Unfortunately however, this metric requires the ability to accurately estimate log probabilities of the model. This is unlikely to be possible for our circuits (or indeed most circuit families), and so we can only use this evaluation metric when the number of qubits is small enough to allow for tractable computation of probabilities. 

Luckily for us, the maximum mean discrepancy with respect to a test set is also a relatively common metric for model evaluation (\cite{o2021evaluation, lueckmann2021benchmarking, sutherland2016generative}) and was found by \citet{xu2018empirical} to have particularly appealing properties compared to other methods. 
\begin{definition}[MMD$^2$ with respect to a test set]
    Given a generative model $q_{\btheta}$ the {\normalfont{MMD$^2$}} with respect to a test set $\mathcal{X}_{\text{test}}=\{\x_i\}$ is 
    \begin{align}
         {\normalfont{\text{MMD}}}^2({P}_{\mathcal{X}_{\text{test}}}, q_{\btheta})
    \end{align}
    where ${P}_{\mathcal{X}_{\text{test}}}(\x)=\frac{1}{\vert\mathcal{X}_{\text{test}}\vert}\sum_{\x_i\in \mathcal{X}_{\text{test}}}\mathbb{I}(\x_i=\x)$ is the empirical distribution of the test data.
\end{definition}
In the above $\mathbb{I}(\cdot)$ is the indicator function that returns 1 if the condition is met and zero otherwise. Since we have shown how to estimate the MMD$^2$ classically, we can use this metric to evaluate parameterised IQP models with large numbers of qubits (i.e. taking $\mathcal{X}=\mathcal{X}_{\text{test}}$ in \eqref{eq:mmd_beast} provides an unbiased estimate). If we are able to sample from the generative model, \eqref{eq:MMD_greton} can be used instead. The MMD is a distance metric, so lower values imply better performance.

We also make use of a test recently proposed in \cite{ravuri2023understandingdeepgenerativemodels}, called the Kernel Generalized Empirical Likelihood (KGEL). This test serves as a diagnostic tool that can identify mode dropping and mode imbalance in trained generative models, and is the solution to the following convex optimisation problem\footnote{In the publication \cite{ravuri2023understandingdeepgenerativemodels}, an additional feature map is applied to the data, which can be used to map high dimensional data to a more informative subspace. It is not clear if this can be combined with our techniques for IQP circuits however, so we take the feature map to be the identity here.}. 
\begin{definition}[Kernel Generalized Empirical Likelihood]
    The {\normalfont{KGEL}} with respect to a test set $\mathcal{X}_{\text{test}}=\{\x_i\}$ and a set of witness points $\{\t_j\}$ is the solution to the following convex optimisation problem:
    \begin{align}\nonumber
    \text{\normalfont{KGEL}}(\mathcal{X}_{\text{test}}, q_{\btheta}) = 
    \min_{\{\pi_i\}} D_{KL}(P_{\boldsymbol{\pi}}\vert\vert P_{\mathcal{X}_{\text{test}}}) \\[3pt] \mbox{\normalfont{subject to}} \quad
    \sum_{i=1}^n \pi_i 
    {\scriptscriptstyle \begin{bmatrix}
    k(\x_i, \t_1)\\
    \vdots \\
    k(\x_i, \t_W)
    \end{bmatrix}}
    =
    \mathbb{E}_{\y\sim q_{\btheta}} \begin{bmatrix}
    k(\y, \t_1)\\
    \vdots \\
    k(\y, \t_W)
    \end{bmatrix}. \label{eq:kgel}
    \end{align}
\end{definition}
Here $P_{\boldsymbol{\pi}}(\x)=\sum_{i=1}\pi_i\mathbb{I}(\x_i=\x)$ (with $\pi_i>0, \sum_i \pi_i=1$), $D_{KL}$ is the Kullback–Leibler divergence (or relative entropy), and the points $\boldsymbol{t}_i$ are called witness points that (like the points in $\mathcal{X}_{\text{test}}$) are typically sampled from the ground truth distribution $p$. The right hand side of \eqref{eq:kgel} can therefore be understood as a vector of expected distances to each of the witness points under the kernel $k$, which here we take to be the Gaussian kernel with a specified bandwidth. 
If the generative model $q_{\btheta}$ has collapsed to a single mode (say close to the point $\t_1$), then in order to satisfy the constraint \eqref{eq:kgel}, the probabilities $\pi_i$ will be weighted heavily towards those test points $\x_i$ that belong to the same mode. This results in a larger value of the KL divergence with respect to the empirical distribution $P_{\mathcal{X}_{\text{test}}}$, but more importantly, the mode dropping can be diagnosed by inspecting the solution $\{\pi_i\}$ returned by the convex optimisation solver. We show in App.~\ref{app:kgel} that the right hand side of \eqref{eq:kgel} can be approximated efficiently with a classical algorithm when $q_{\btheta}$ is a parameterized IQP circuit. As a result, we can evaluate mode dropping of large quantum generative models without the need to sample. Like the MMD$^2$, the KGEL is implemented in the IQPopt package (\cite{iqpopt}), which we use to investigate mode imbalance of our models in the following section.

Finally, since we have the ability to estimate expectation values of parameterised IQP circuits we can also estimate the covariance between the $i^{th}$ and $j^{th}$ elements of $\x$.
\begin{definition}[covariance]
The covariance between the $i^{th}$ and $j^{th}$ elements of $\x$ for the generative model $q_{\btheta}(\x)$ is
\begin{align}
    {\normalfont{cov}}(x_i,x_j) = \langle Z_{i}Z_j \rangle_{q_{\btheta}} - \langle Z_{i} \rangle_{q_{\btheta}}\langle Z_{j} \rangle_{q_{\btheta}}.
\end{align}
\end{definition}
Using this we can construct the covariance matrix of the trained model in order to visually inspect the two body correlations and compare this to the true distribution. 

\section{Experiments}\label{sec:experiments}
In this section we apply the theoretical work of the previous sections to a number of numerical experiments that investigate the potential of parameterised IQP circuits to serve as useful generative models. The code that was used to produce these results can be found at \href{https://github.com/XanaduAI/scaling-gqml}{\texttt{github.com/XanaduAI/scaling-gqml}}, along with scripts to generate or download the datasets. We focus on the following six datasets. 
\begin{enumerate}
    \item \emph{2D Ising dataset}: A dataset of bitstrings of length 16, that are sampled from a 16-spin classical Ising distribution with a square lattice Hamiltonian. 
    \item \emph{Binary blobs dataset}: A dataset of bitstrings of length 16 that are sampled close in Hamming distance to one of eight pre-specified patterns.
    \item \emph{D-Wave dataset}: A dataset of bitstrings of length 484 that are sampled from quenching 484-qubits in a D-Wave Advantage system with a Pegasus lattice topology, taken from \cite{scriva2023accelerating}. 
    \item \emph{Binarized MNIST dataset}: A dataset of bitstrings of length 784 constructed by binarizing the full-pixel MNIST handwritten digits dataset.
    \item \emph{Scale free network dataset}: A dataset of bitstrings of length 1000 sampled from a classical Ising distribution on a 1000-spin system whose Hamiltonian connectivity corresponds to a scale-free network. 
    \item \emph{Genomic dataset}: A real-world genomic dataset of bitstrings of length 805 that correspond to single nucleotide polymorphisms of a highly variable section of the human genome of length 805, taken from \cite{yelmen2021creating}. 
\end{enumerate} 
As well as training parameterized IQP circuits, we also train two energy-based classical generative models as well as the stochastic bitflip model of Sec.~\ref{sec:bitflip} to serve as comparisons. We thus consider a total of four models:

\begin{itemize}
    \item \emph{IQP model}: A quantum generative model corresponding to a parameterised IQP circuit.
    \item \emph{Bitflip model}: A stochastic bitflip model, as described in Sec.~\ref{sec:bitflip}. 
    \item \emph{RBM model}: A restricted Boltzmann machine, implemented via sci-kit learn's \texttt{BernoulliRBM} class.
    \item \emph{EBM model}: An energy based model whose energy function is given by a feedforward neural network. 
\end{itemize}
Compute intensive operations were performed with the aid of the digital research alliance of Canada's Niagara cluster (40 core, 202GB RAM per node) or an in-house server featuring NVIDIA's Grace Hopper G200 superchip (72 core, 480GB + H100 GPU). All calculations were performed on CPU, with the exception of the EBM training for the D-Wave dataset, which used the Grace Hopper GPU. 

\subsection{Training strategy for the IQP model}\label{sec:trainstrat}
Here we describe the specific training strategy we adopted to train the IQP model. An identical strategy was used to train the bitflip model.

\subsubsection{Choice of loss function}
The parameterised IQP circuits are trained via the squared maximum mean discrepancy loss of \eqref{eq:expmmd}. We use the average of a number of MMD$^2$ values for different values of the bandwidth $\sigma$. That is, our loss takes the form 
\begin{align}
    \mathcal{L} = \frac{1}{L}\sum_{i=1}^{L} \text{MMD}_{\sigma_i}^2(P_{\mathcal{X}_{\text{train}}}, q_{\btheta})
\end{align}
for a choice $\{\sigma_1, \cdots,\sigma_{L}\}$ of bandwidths and $P_{\mathcal{X}_{\text{train}}}$ the empirical training distribution. For experiments 1 and 2 (which involve 16 qubit models), the specific choice $\{\sigma_1, \sigma_2\}=\{0.6, 0.3\}$ was used, which corresponds to sampling observables $Z_{\a}$ in \eqref{eq:expmmd} with an average Pauli weight of $2$ and $6$ respectively. For all other experiments, we used a choice of three bandwidths $\{\sigma_1, \sigma_2, \sigma_3\}$, where $\sigma_1$ is such that the average Pauli weight of $Z_{\a}$ is $2$, $\sigma_3$ is the square root of the median heuristic, and $\sigma_2$ is given by
    $\sigma_2 = \sqrt{(\sigma_1^2 + \sigma_3^2)/2}$.
The values and implied average Pauli weights of $Z_{\a}$ for each experiment are shown in Table~\ref{tab:weights}, which can be used to understand the order of correlations probed by each bandwidth. In practice, it can be beneficial to consider several bandwidths (as was done in \cite{li2015generative}), since gradient information may only be possible for low body correlations at the start of training (\cite{rudolph2024trainability}); terms in $\mathcal{L}$ focusing on higher order correlations therefore become more significant the model is trained. 

\begin{table}
\centering
\begin{tabular}{|c|c|c|c|c|}
\hline
& $n$ & $\sigma_1$ & $\sigma_2$ & $\sigma_3$ \\ \hline
2D Ising & 16 & 1.3 (2) & 0.6 (6) & n/a \\ \hline
Binary Blobs & 16 & 1.3 (2) & 0.6 (6) & n/a \\ \hline
D-wave & 484 & 7.8 (2) & 6.1 (4) & 3.9 (8) \\ \hline
MNIST & 784 & 9.9 (2) & 7.4 (4) & 3.4 (17) \\ \hline
Scale free & 1000 & 11.2 (2) & 8.3 (4) & 3.8 (17) \\ \hline
Genomic & 805 & 10.0 (2) & 7.7 (4) & 4.2 (11) \\\hline
\end{tabular}
\caption{The number of qubits of the IQP model (column $n$) and the values of the bandwidths used for training and evaluation. The parentheses show the corresponding average operator weights (the average Pauli weight of $Z_{\a}$ in \eqref{eq:expmmd}, rounded to the nearest integer).}
\label{tab:weights}
\end{table}

\subsubsection{Data-dependent parameter initialisation}
We found that a wise choice of parameter initialisation is crucial to obtain good solutions for larger problems. Choosing to initialize all parameters uniformly at random typically leads to a situation where the loss does not decrease, which may due to the presence of barren plateaus in the loss landscape (\cite{mcclean2018barren}) (see Sec.~\ref{sec:discussion} for a more in depth discussion on this issue). To mitigate these issues we adopted a technique in which the magnitude of the initial parameters are dependent on the training data. In particular, for single qubit gates with $X$ generator acting on qubit $j$, we initialize parameters to values $\arcsin(\sqrt{\langle x_j \rangle })$, where $\langle x_j \rangle$ is the mean value of the $j^{th}$ dimension of the training data. This choice ensures that if all other parameters are set to zero, the model is equivalent to a product distribution with the same single qubit marginal distributions as the training data. Parameters of two-qubit gates acting on qubits $j,k$ are initialized proportional to the covariance between the $j^{th}$ and $k^{th}$ dimensions of the training data (the specific constant of proportionality is left as a free hyperparameter), where we convert the training data to $\pm1$ values rather than binary. The logic here is similar: if two features are highly correlated, then it is likely the corresponding parameter will be relatively large since parameterised IQP gates enact a coherent bitflip. All other parameters (if there are any) are initialized via independent zero-mean normal distributions, whose standard deviation is another hyperparameter. 

\subsubsection{Training via stochastic gradient descent}
\begin{figure*}
    \includegraphics[width=\textwidth]{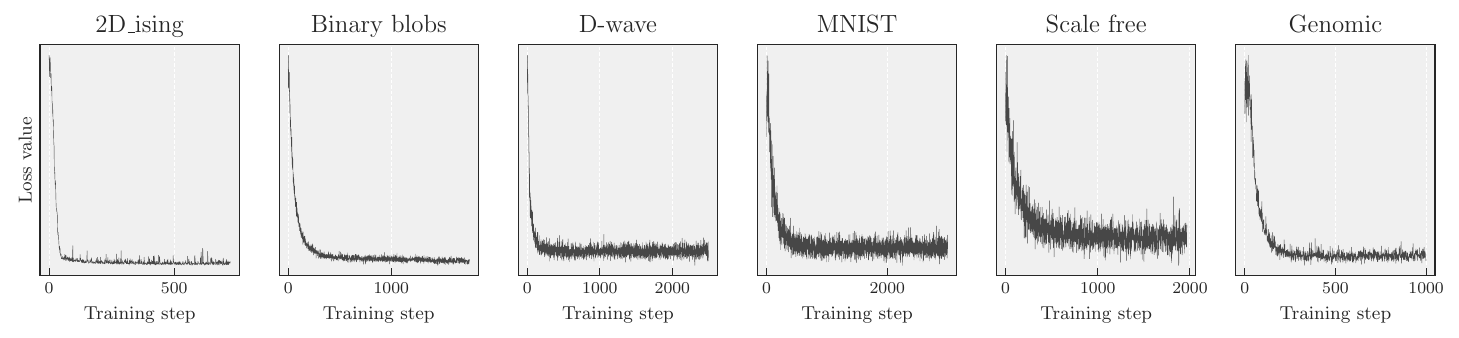}
    \caption{Training loss plots from training the parameterised IQP model on each of the six datasets. \label{fig:lossplots}}
\end{figure*}

For each training update step, we compute an unbiased estimate of the loss via \eqref{eq:mmd_beast}, setting $\vert\mathcal{A}\vert=\vert\mathcal{Z}\vert = 1000$ and taking $\mathcal{X}$ to be the set of training data. Gradients are estimated from this via automatic differentiation in JAX and used via the ADAM update to train the model using the IQPopt package (\cite{iqpopt}), and training is stopped after a predetermined number of steps or when a convergence criterion is met. Training in this way appears to be smooth and convergence is achieved in a relatively small number of steps. In Fig.~\ref{fig:lossplots} we show the loss plots resulting from training the model for each of the above experiments. 

\subsection{Classical generative models}
In this section we describe each of the classical generative models in greater detail.

\subsubsection{Bitflip model}
The bitflip model has an identical structure to the IQP model, differing only in the form of the expectation values as given by \eqref{eq:expval_bitflip}. We therefore use an identical strategy to the IQP model to train this model, which allows for a clean and fair comparison between the two. As with the IQP model, the training was performed via the IQPopt package. 

\subsubsection{Energy based models}
The remaining two classical models are both instances of energy based generative models. These models parameterise probability distributions over bitstrings $\s$ of the form 
\begin{align}
    P_{\btheta}(\s) = \frac{\exp(-E_{\btheta}(\s))}{Z_{\btheta}},
\end{align}
where $E_{\btheta}(\s)$ is the energy function and $Z_{\btheta}=\sum_{\s}(\exp(-E_{\btheta}(\s)))$ is a normalisation constant know as the partition function.
\\

\emph{RBM Model}---For the RBM model, $\s$ contains the data vector $\x$ as well as another binary vector $\h$ whose elements are called hidden neurons. The energy function is equivalent to a classical Ising energy where the Hamiltonian has a bipartite structure between the data and hidden neurons:
\begin{align}
    E_{\btheta}(\s) = E_{\btheta}(\x,\h) = -\boldsymbol{\alpha}\cdot \x - \boldsymbol{\beta}\cdot \h - \x^T W \h,
\end{align}
where $\btheta = (\boldsymbol{\alpha}$, $\boldsymbol{\beta}, W)$ contains the trainable parameters of the model. To obtain the generative model over the data, one marginalizes the distribution over the hidden units:
\begin{align}
    q_{\btheta}(\x) = \sum_{\h}P_{\btheta}(\x,\h) = \sum_{\h}\frac{\exp(-E_{\btheta}(\x,\h))}{Z_{\btheta}}.
\end{align}
 The RBM is trained with persistent contrastive divergence, as implemented by the \texttt{fit} method of sci-kit learn's \texttt{BernoulliRBM} class. Parameter initialisation follows the default behavior of sci-kit learn, which corresponds to a a Xavier initialisation (\cite{glorot2010understanding}). \\

 \emph{EBM Model}---The EBM model has a simpler structure and uses a neural network to parameterise the energy function directly, without the use of additional stochastic neurons. That is, for the EBM we have $\s=\x$ and the function $E_{\btheta}(\x)$ is given by a deterministic feedforward neural network with a specified number of hidden layers with corresponding parameters $\btheta$. The EBM model is trained via the standard contrastive divergence algorithm, which we implement in JAX (\cite{jax}) and Flax (\cite{flax2020github}). Parameter initialisation follows the default behavior of Flax, which corresponds to a LeCun normal initialisation (\cite{klambauer2017self}). \\

To generate samples from both models we need to use Markov chain Monte Carlo methods to attempt to sample from the distribution $P_{\btheta}(\s)$. This is known to be computationally expensive, since the typical length of the Markov chain needed to reach the equilibrium distribution is generally unknown and can be long, especially for energy functions that contain deep local minima. For the RBM model, the bipartite structure of the energy function allows for a Gibbs sampling procedure to update the all data or hidden neurons in parallel. For the EBM model, this is generally not possible, and so bits are updated individually using the standard Metropolis Hastings algorithm, which therefore requires more computational effort. In order to obtain the highest quality samples for evaluation, we seed an independent Markov chain for every sample, selecting the last bitstring after a large number of MCMC steps (see Table \ref{tab:steps} for precise values). The two models were implemented via the \texttt{RestrictedBoltzmannMachine} and \texttt{DeepEBM} classes of the \texttt{qml-benchmarks} package (\href{https://github.com/XanaduAI/qml-benchmarks}{github.com/XanaduAI/qml-benchmarks}). 

\begin{table}
\centering
\begin{tabular}{|p{7.5cm}|p{2.3cm}|p{3.4cm}|}
\hline
\textbf{Hyperparameter (IQP/Bitflip model)} & \textbf{Experiments} & \textbf{Range} \\ \hline
Largest Pauli weight of gate generators  & 1,2 & [2,4,6] \;(2)  \\ \hline
Number of ancilla qubits  & 1,2 & [0,8]\; (0) \\ \hline
Initial learning rate & 1,2,3,4,5,6 & 0.0001 - 0.1 \\ \hline
Parameter initialisation scale factor (two-qubit gates) & 1,2,3,4,5,6 & 0.0001 - 1.0 \\ \hline
Parameter initialisation scale factor (>2 qubit gates) & 1,2 & [0, 0.0001] \; (0) \\\hline 
\end{tabular}

\vspace{0.2cm} 

\begin{tabular}{|p{7.5cm}|p{2.3cm}|p{3.4cm}|}
\hline
\textbf{Hyperparameter (RBM model)} & \textbf{Experiments} & \textbf{Range} \\ \hline
Number of hidden units  & 1,2,3,5 & 4-1024  \\ \hline
Learning rate  & 1,2,3,5 & 0.00001 - 0.01 \\ \hline
Batch size & 1,2,3,5 & 16-64 \\ \hline
\end{tabular}

\vspace{0.2cm} 

\begin{tabular}{|p{7.5cm}|p{2.3cm}|p{3.4cm}|}
\hline
\textbf{Hyperparameter (EBM model)} & \textbf{Experiments} & \textbf{Range } \\ \hline
Neural network layer structure  & 1,2,3,5 & variable \\ \hline
Initial learning rate  & 1,2,3,5& \small{[0.00001, 0.001, 0.001]} \\ \hline
Contrastive divergence steps & 1,2,3,5 & [1,10] \\ \hline
Batch size & 1,2,3,5 & 16-128 \\ \hline
\end{tabular}

\caption{Hyperparameters used in a preliminary grid search for each model. The experiments column shows for which of the six experiments the hyperparameter was varied. In the case that all the specified experiments searched over the same values the range is stated as a list; otherwise the range shown is the total range over all experiments in which the hyperparameter was varied (for precise values for each experiment can be found in \href{https://github.com/XanaduAI/scaling-gqml/blob/b907bb119c45ee85c87f1eb91867f4a7d281f5be/paper/hyperparam_opt/hyperparameter_settings.py}{the accompanying repository}). The values shown in rounded parentheses correspond to the default values that were used if the hyperparameter was not varied.}
\label{tab:hpopt}
\end{table}

\begin{table}
  \centering
  \resizebox{\textwidth}{!}{%
  \begin{tabular}{|p{1.1cm}|c|c|c|c|c|c|c|c|c|c|c|}
    \hline
    \multirow{2}{3cm}{\textbf{Model}} & \multicolumn{2}{c|}{\textbf{Ising}} & \multicolumn{2}{c|}{\textbf{Blobs}} & \multicolumn{2}{c|}{\textbf{D-Wave}} &
    \multicolumn{2}{c|}{\textbf{MNIST}} & 
    \multicolumn{2}{c|}{\textbf{scale free}}  \\
    \cline{2-11}
    & \textbf{train} & \textbf{sample}  &  \textbf{train} & \textbf{sample} & \textbf{train} & \textbf{sample} &
    \textbf{train} & \textbf{sample} &
    \textbf{train} & \textbf{sample}  \\
    \hline
    IQP & 700 & exact & 1750 & exact & 2500 & n/a & 3000 & n/a & n/a & n/a  \\ \hline
    Bitflip & 2500 & exact & 2500 & exact & 2500 & exact & 3000 & exact & 2600 & exact \\ \hline
    RBM & 10k & 3200 & 10k & 3200 & 5000 & 300k & n/a & n/a & 5000 & 150k   \\ \hline
    EBM & 500k & 64k  & 500k & 64k & 5m & 4.5m & n/a & n/a & 2m & 1m  \\ \hline
  \end{tabular}
  }
  \caption{Training and sampling steps for experiments 1 to 5 (k $=10^3$, m $=10^6$). For all models except RBM, the value for train is the number of gradient steps to train the model. For RBM, it is the number of epochs of training (so the number of steps is much larger). For EBM, the sample column shows the number of MCMC steps to obtain each sample used for evaluation. For RBM, it is the number of Gibbs steps between the data and hidden units of the model to obtain each sample. For the bitflip model and IQP model  (below 16 qubits) sampling can be achieved exactly without the need for MCMC. For the genomic dataset (not shown here), both the IQP and bitflip models were trained for 1000 steps. \label{tab:steps}}
\end{table}

\begin{table}
  \centering
  \begin{tabular}{|p{1.1cm}|c|c|c|c|c|c|}
    \hline
      & \textbf{Ising} & \textbf{Blobs}  &  \textbf{D-Wave} & \textbf{MNIST} & \textbf{Scale free} & \textbf{Genomic}  \\
    \hline
    IQP & 2516 & 14892 & 117370 & 307720 & 500500 & 324415 \\ \hline
    Bitflip & 14892 & 2516 & 117370 & 307720 & 3991 & 324415  \\ \hline
    RBM & 4368 & 1104 & 497124 & n/a & 251250 & n/a  \\ \hline
    EBM & 1800 & 348  & 235224 & n/a & 484 & n/a  \\ \hline
  \end{tabular}
  \caption{Number of trainable parameters in each of the models we trained, as selected by the hyperparameter grid search. } \label{tab:params}
\end{table}

\subsection{Hyperparameter optimisation pipeline}
For each dataset and model, a hyperparameter grid search was performed to identify the most promising choice of hyperparameters on which to train the models for evaluation. As is always the case, the specific choice of hyperparameter grid, as well as the relative effort and computational budget spent on each model can greatly affect the final results (\cite{bowles2024better}). We strove to be as fair as possible in our experiments, and put a comparable effort into both training and searching hyperparameters for each model. For high dimensional datasets, limits of the hyperparameter search space and the total training time were often dictated by computational requirements, i.e.\ by the job time of single node jobs on the compute clusters. In Table~\ref{tab:hpopt} we detail the different hyperparameters that were searched over for each model. 

For each choice of hyperparameters, the corresponding model was trained and evaluated on a single train/validation split of the training data. We did not perform multiple cross validations, opting to use the available computational resources to enlarge the search grid rather than perform multiple training runs (and our initial investigations suggested performance was largely independent of the particular split). The validation set was used to evaluate the hyperparameter choice: to select the best hyperparameters, the average MMD$^2$ across a number of bandwidths (the same ones as shown in Table~\ref{tab:weights}) was estimated, and the hyperparameter choice with the lowest value was selected. These hyperparameters were then used to train the model again, this time using the full training data. Typically, the number of training steps was significantly increased in this final training to achieve the best possible results. 

The hyperparameter grids, choice of best parameters, loss plots from training, trained parameters, as well as instructions on how to load and sample from the trained models can be found in the accompanying code repository. The total steps used to train and sample from the models is shown in Table \ref{tab:steps}, and the number of parameters of the models as selected by the grid search are shown in Table \ref{tab:params}.

\subsection{Dataset descriptions and results}
In this section we describe the datasets in further detail and interpret the obtained results. Scripts to generate or download each of the datasets can be found in the datasets directory of the accompanying repository. The synthetic data for experiments 1,2 and 5 was generated using the \texttt{ising} (experiments 1 and 5) and \texttt{spin\_blobs} functions (experiment 2) of the \emph{qml-benchmarks} package respectively (\href{https://github.com/XanaduAI/qml-benchmarks}{github.com/XanaduAI/qml-benchmarks}). 

\subsection*{Experiment 1: 2D Ising dataset}


This dataset corresponds to a thermal distribution at temperature $T= 3/k_B$ of an Ising spin system on a $4 \times 4$ square lattice with periodic boundary conditions. The coupling weights of the Hamiltonian are sampled independently as uniform random numbers in the range $[0,2]$, and there are no local bias terms. Since Ising distributions without local biases are invariant with respect to flipping all bits, the distribution satisfies the spin flip symmetry $p(\x)=p(\bar{\x})$ described in Sec.~\ref{sec:sym}, and we therefore train the IQP and bitflip models with this symmetry enforced. We generated a total of 800000 configurations by performing Metropolis Hastings Monte Carlo sampling on 8 independent Markov chains, from which we randomly selected 5000 points to form a training set, and 50000 points to form a test set from equally spaced points on the chains. 

\begin{figure*}
    \centering
    \includegraphics[width=1.0\textwidth]{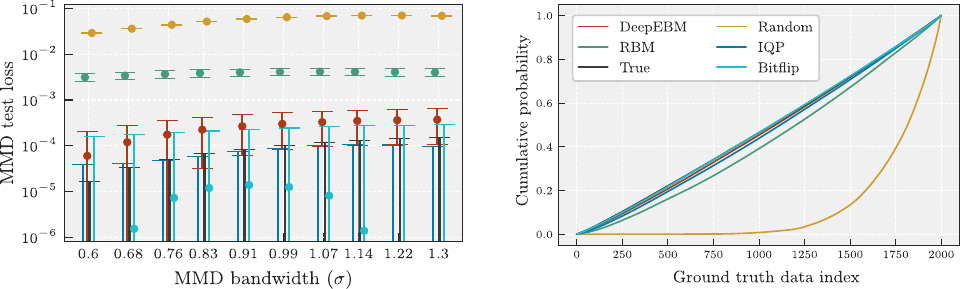}
    \caption{(left) The squared maximum mean discrepancy evaluated on a test set for each of the models for the 2D Ising data. Error bars denote one standard deviation. (right) The cumulative probability distribution returned by the KGEL test.}
    \label{fig:ising}
\end{figure*}

\begin{figure*}
    \centering
    \includegraphics[width=0.9\textwidth]{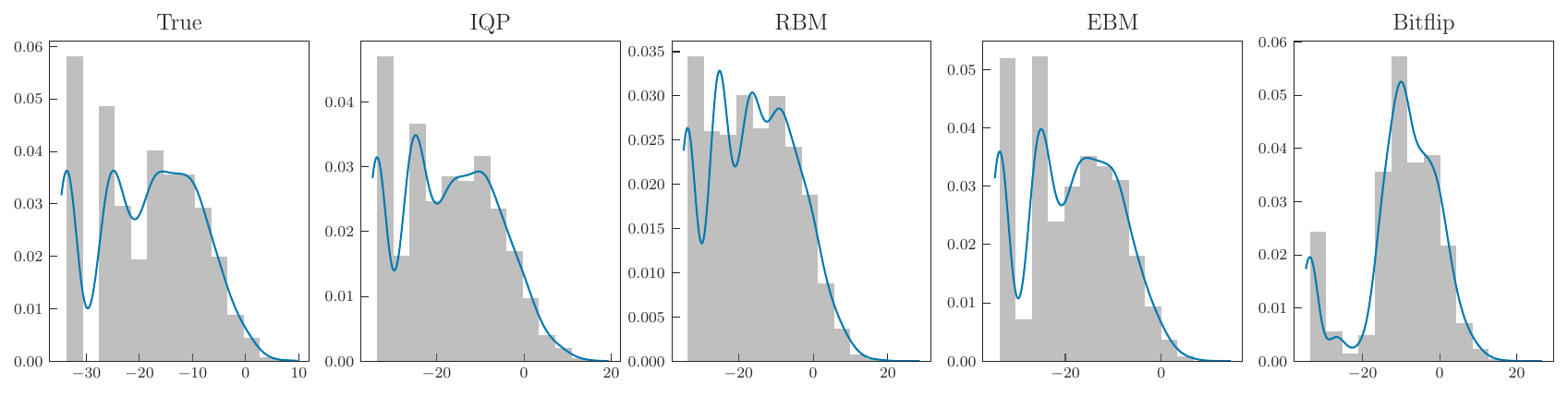}
    \includegraphics[width=0.9\textwidth]{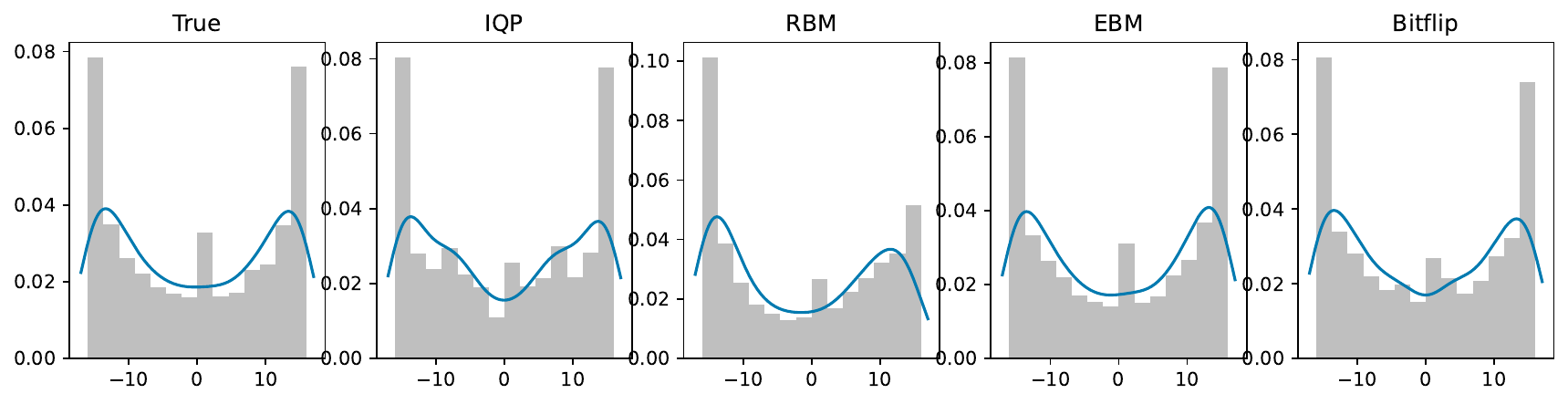}
    \caption{Distributions computed through kernel density estimation (blue curves) of the Ising energy (top) and magnetisation (bottom) for the true distribution and each of the trained models.}
    \label{fig:ising_dists}
\end{figure*}

In Fig.~\ref{fig:ising} (left) we show the MMD$^2$ evaluated on the test data over the range of bandwidths on which the IQP model was trained. Both the IQP and bitflip models perform very well, resulting in MMD values equal to zero to within 1 standard deviation error. The reason for the superior performance relative to the EBM may be partly due to the fact that unlike the EBM model, the IQP and bitflip models were trained with the MMD loss function: if one calculates the log likelihood of a test set\footnote{calculation of the log likelihood for the RBM model is computational intractable, and not implemented for the bitflip model.}, one finds the EBM ($-7.53$) performs slightly better than the IQP model ($-7.82$). Since the IQP and Bitflip models were trained with the spin flip symmetry of Sec.~\ref{sec:sym} enforced, they also possess a relevant inductive bias not present in the other models, which likely contributed to the good performance as well. The RBM model performs relatively poorly in comparison, which we suspect is due to mode imbalance: in Fig.~\ref{fig:ising_dists} we see that the distribution of magnetisation is asymmetric, with a preference for negatively aligned spins. This is supported by the corresponding KGEL distribution of Fig.~\ref{fig:ising} (right), which is further from a uniform distribution (corresponding to a straight line) than other models. The similar performance of the bitflip and IQP models suggests that coherence is playing a small role in this experiment, although the bitflip model does appear to struggle to sample low energy configurations. 

\subsection*{Experiment 2: Binary blobs dataset}
\begin{figure*}
    \centering
    \includegraphics[width=1.0\textwidth]{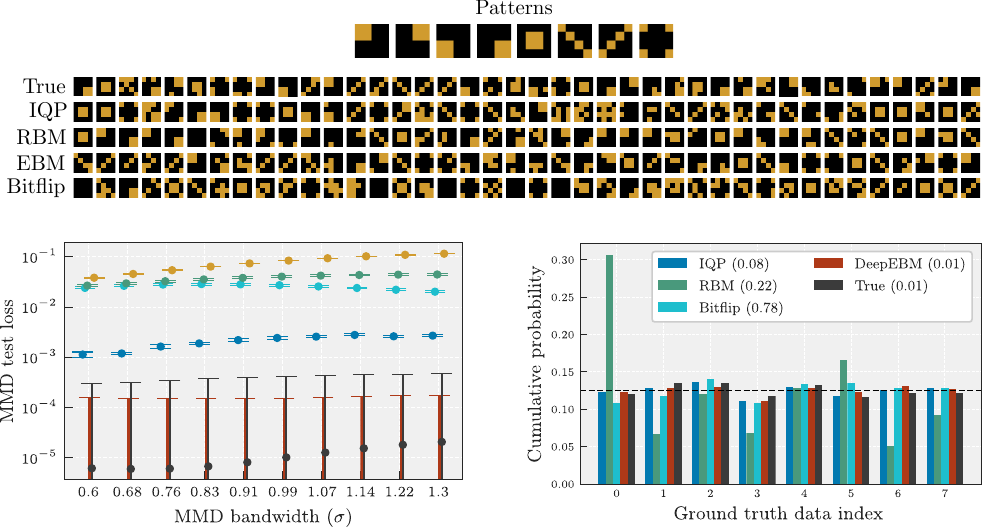}
    \caption{(top) The eight base images used to construct the binary blobs dataset, and samples from the ground truth distribution and each of the trained models. Dark squares indicate a 1 in the corresponding bitstring. (bottom) The test MMD$^2$ values (left) and the KGEL distributions for the trained models (right). The gold points in the MMD plot correspond to randomly sampled data. For the KGEL distributions, the probability distribution $\pi_i$ returned from the convex optimisation has been binned according to the eight modes so that mode imbalance can be diagnosed.}
    \label{fig:8blobs}
\end{figure*}

This dataset was constructed as a bitstring analog to the commonly used `Gaussian blobs' datasets that are comprised of real vectors. To generate the data, one of eight specified bitstrings of length 16  (shown at the top of Fig.~\ref{fig:8blobs}) was randomly chosen and each pixel was independently flipped with probability $0.05$. This  creates a distribution over bitstrings that features eight clearly separated modes that can be visually identified. This process was used to create train and test datasets of sizes $5000$ and $10000$; samples from this distribution can be seen in the `True' row at the top of Fig.~\ref{fig:8blobs}. 


As for the 2D Ising data, the RBM model faces issues with mode imbalance, resulting in poor values of the test MMD$^2$. This can be seen from the results of the KGEL test, where there is a clear preference to sample the first configuration (which can also be seen from the generated samples above). The EBM performs exceptionally well here, achieving results that are indistinguishable from the test data. The IQP model is able to capture all of the patterns and does not appear to suffer from extreme mode imbalance, however sometimes produces configurations that are far from the ground truth distribution. This can be seen from the fourth example configuration from Fig.~\ref{fig:8blobs} that has a Hamming weight of 11. This is despite the model using all gates acting on six qubits or less, resulting in a total of 14892 parameters (significantly more than the EBM's 1700 parameters). The superior performance of the EBM is further supported by a higher log likelihood score of -5.34, compared to -6.35 for the IQP model. The bitflip model performed badly, often producing bitstings far from the ground truth.  The MMD$^2$ values of the bitflip model are better than those of the RBM, despite the RBM's samples being arguably more visually pleasing. We suspect that this is because the MMD$^2$ metric does not only reflect how visually pleasing a sample is, but is dependent on other factors such as mode imbalance. Indeed, in the machine learning literature, even evaluation metrics designed to select visually pleasing samples can still lead to strange results (\cite{barratt2018note}).

\begin{figure*}
    \centering
    \includegraphics[width=1.0\textwidth]{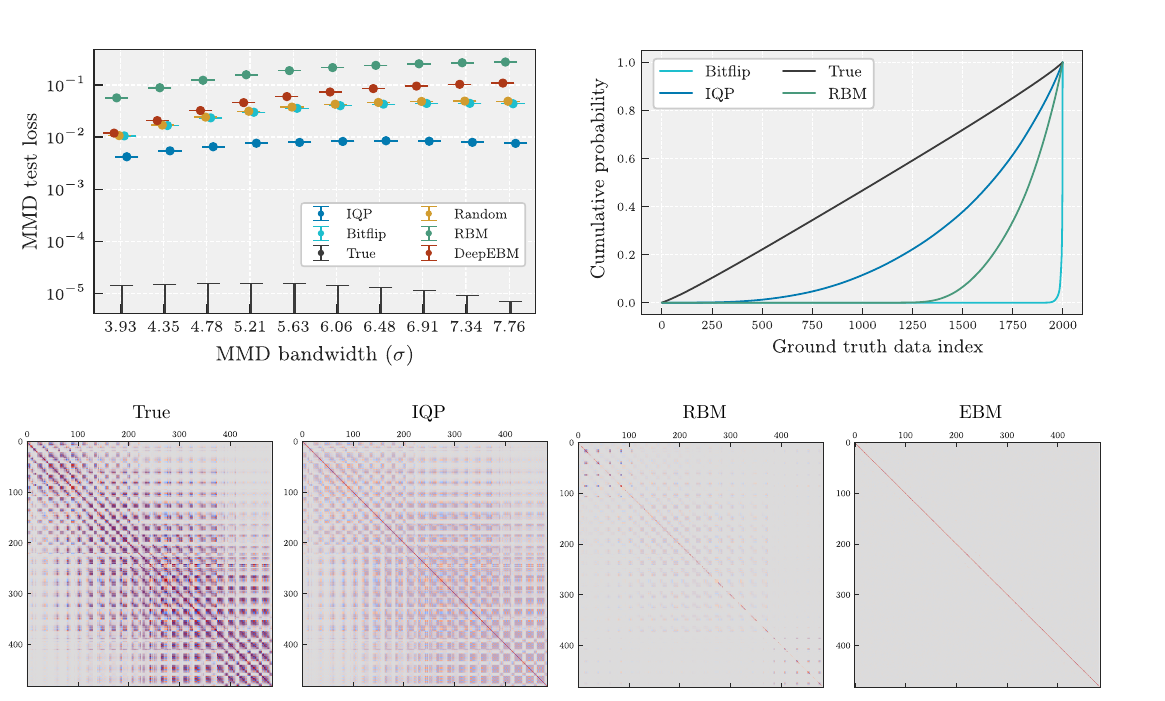}
    \caption{(top): Values of the squared maximum mean discrepancy of the trained models evaluated with respect to a test set (left), and the cumulative probability distributions returned by the KGEL test (for $\sigma=3.93$ and 10 witness points). The KGEL result for the EBM is not shown since the solver failed to produce a solution.}
    \label{fig:dwave}
\end{figure*}

\subsection*{Experiment 3: D-Wave dataset}\label{sec:exp:sf}
This dataset was generated from spin configurations sampled from a \href{https://www.dwavesys.com/solutions-and-products/systems/}{D-Wave Advantage processor}. The dataset appeared in \citet{scriva2023accelerating}, where it was used to generate configurations to seed and accelerate Markov chain Monte Carlo methods to sample from thermal distributions of a spin glass model at low temperatures. We focus on data that was obtained by annealing a 484-qubit spin system coupled via D-Wave's Pegasus graph topology, that underwent quantum annealing for $100\mu s$. The training and test data consists of 10000 points and 60000 points respectively, taken from the data available from the paper's code repository. This results in a  dataset which features correlations over relatively large distances, which can be seen from the covariance matrix plot of the data shown in Fig.~\ref{fig:dwave}. Both the IQP and bitflip models were trained using all two-qubit and single-qubit gates on the 484 qubits of the model, resulting in models with over 100000 parameters. 

Interestingly, although no model managed to capture the distribution very well, the IQP model excelled at this problem compared to the classical models. From the covariance matrix plot of Fig.~\ref{fig:dwave} one sees that the overall structure of the two body correlations have been well captured, however are significantly weaker than those of the true distribution. The RBM model has likely suffered issues of mode collapse since many of the diagonal elements of the covariance matrix (i.e.\ the variances of each bit of the distribution) are close to zero, meaning some bits are effectively constant. Despite intense training and sampling, the EBM model produced poor results, with only some faint short range correlations visible from the covariance matrix. Both the RBM and EBM models scored worse than a random distribution for the test MMD$^2$. We remark that this is not a contradiction, since the models are trained on a different metric which doesn't necessarily minimize the test MMD$^2$, and so unsuccessful training and extreme mode imbalance may result in such values. The bitflip model failed to train significantly and produced a covariance matrix that was indistinguishable from the random distribution, which we do not show. 

\begin{figure*}
    \centering
    \includegraphics[width=1.0\textwidth]{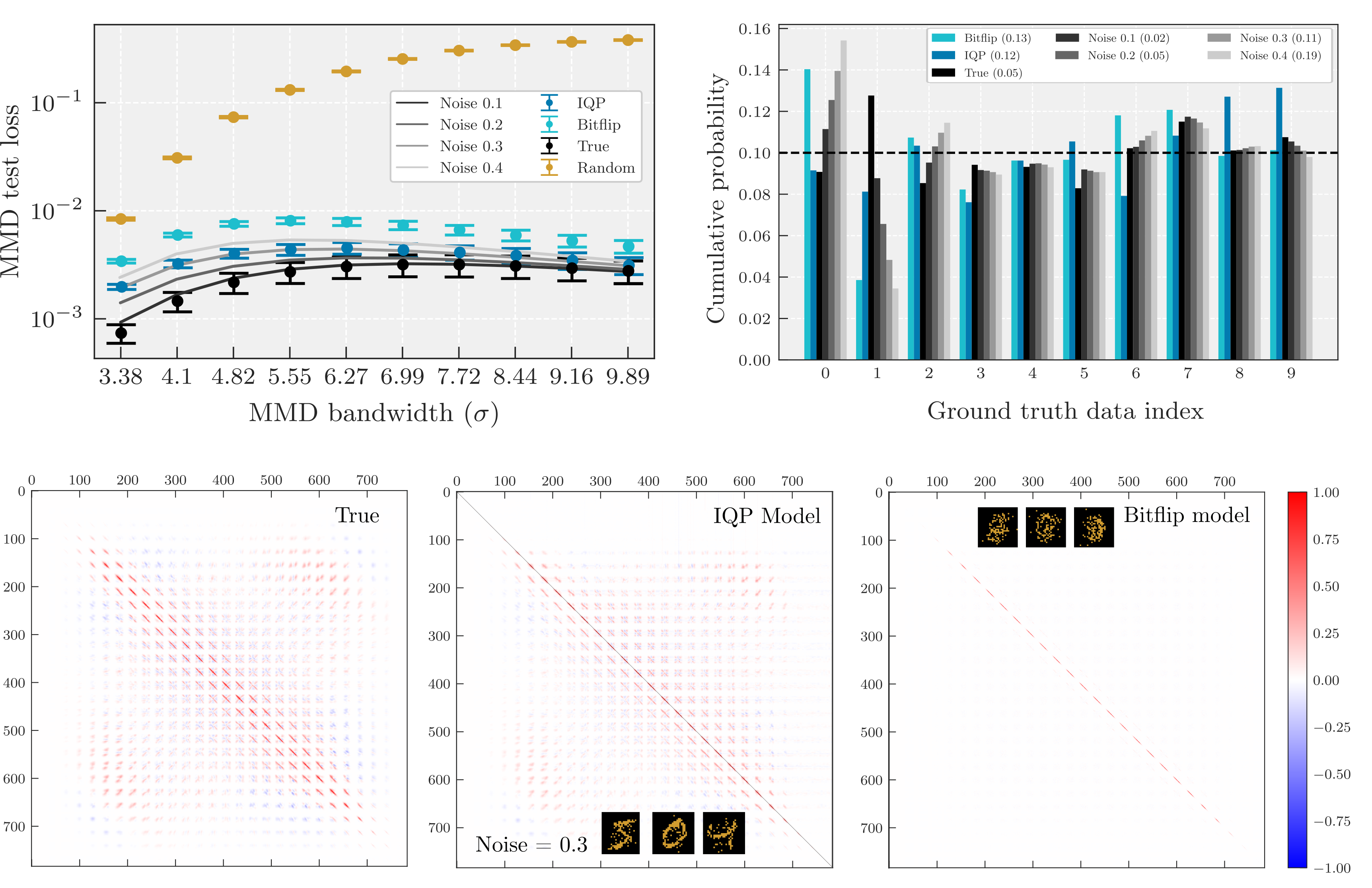}
    \caption{Top left: The test MMD$^2$ values for the trained IQP and bitflip models, uniform random data and data obtained by adding different levels of noise to the training set (see main text for the noise model). Top right: The cumulative probability distribution returned by the KGEL test, binning the probabilities according to digit number. The values of the KGEL objective function are shown in parentheses next to the labels. Bottom: The covariance matrix plots for each model. For the IQP model (centre), the covariances are contrasted to the convariance matrix of noisy data with noise parameter 0.3, and the images are samples from the corresponding models (sampling from the IQP model is not possible).}
    \label{fig:mnist}
\end{figure*}

\subsection*{Experiment 4: Binarized MNIST digits}
The MNIST handwritten digits dataset is probably the most famous dataset in the machine learning literature and is often used as a sanity check to test if a model is performing well. We convert the standard dataset to binary images by thresholding the pixel values, and then flatten the images, resulting in a dataset of bitstrings of length 784. We use the standard train/test split that consists of 50000 training points and 10000 test points. Given that the MNIST dataset has been extensively trained in the machine learning literature, we trained only the IQP and bitflip models on this dataset. The corresponding circuits have 784 qubits and the gate sets consist of all-to-all two-qubit gates, resulting in models with 307720 trainable parameters.  

The results suggest that the IQP model has been able to learn a lot of the structure in the dataset. The clearest evidence for this is in the covariance matrix plot, which mirrors the general structure of the true distribution albeit with slightly weaker correlations. We also compare the results to four noisy distributions. A noisy distribution with noise parameter $p$ corresponds to the following procedure: (i) sampling a point $\x$ from the ground truth distribution then (ii) for each pixel in $\x$, set the pixel to $0$ with probability $p\cdot p_0$,  to $1$ with probability $p\cdot (1-p_0)$, and otherwise leave the pixel unchanged, where $p_0$ is the probability that that pixel takes the value zero over all points in the training data. For $p=1$ the noise distribution is therefore equal to a a product Bernoulli distribution where each pixel is sampled according to its average value. From the test MMD$^2$ plots, the IQP model scores similarly to data with $p=0.3$. In the bottom left of the central covariance matrix plot, we show the covariances for this noisy data, as well as some typical images. The results suggest the quantum model is capable of producing images with enough structure to clearly distinguish the digits by eye, although one would need to sample the quantum circuit to confirm this. The bitflip model fails to learn a lot of structure, as can be seen from the sampled images in the corresponding covariance matrix plot and poor test MMD$^2$ scores. 

To plot the KGEL distribution, we take 10 witness points, where each witness point corresponds to a unique digit from the test set. The plot suggests that no mode has been severely dropped, since the magnitude of variations from the uniform probability are of the same order as those from the true distribution.

\begin{figure}
    \centering
    \includegraphics[width=1\textwidth]{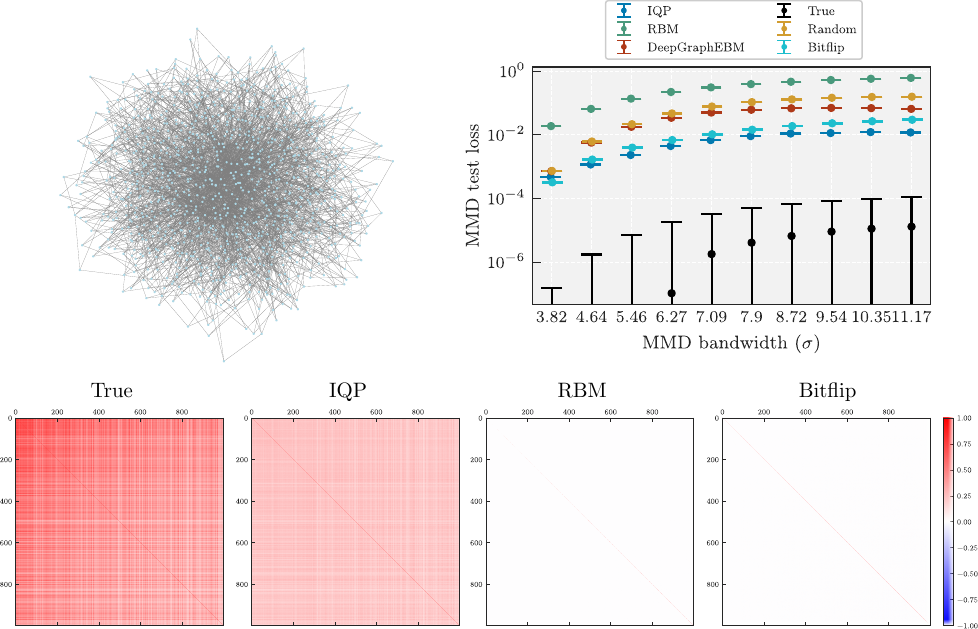}
    \caption{Top left: The scale free graph used to construct the dataset. Top right: The test MMD scores for the scale free dataset. Bottom: The covariance matrices for the true distribution and IQP, RBM and bitflip models.}
    \label{fig:scalefree}
\end{figure}

\subsection*{Experiment 5: Scale free dataset}
This dataset corresponds to a thermal distribution at temperature 1 of a 1000 spin Ising system. The graph describing the two-body interactions of the Ising Hamiltonian corresponds to a scale free network, which is constructed via the Barabasi-Albert algorithm (\cite{albert2002statistical}) with connectivity parameter 2 (see Fig.~\ref{fig:scalefree} for a plot). The Ising energy has a local bias that is dependent on the degree of the corresponding node, which biases the values of that bit towards zero. This gives a crude statistical model of a 1000 person social network in which a value of $0$ indicates a user is active: the bias ensures users with more connections are more active, and users that are connected are more likely to be active at the same time. The data was generated via a Metropolis Hastings algorithm by sampling a million configurations on eight independent Markov chains, and then selecting 20000 train and test points randomly from equally spaced points on the chain. The result is a data set which features positive correlations only, as can be seen from the covariance plot of Fig.~\ref{fig:scalefree}. 

The gate sets for the IQP and bitflip models were first taken to reflect the structure of the graph: for the initial grid search, gate sets with two-qubit gates between either nearest neighbors, or nearest and next nearest neighbors on the graph were used. Only the IQP model produced results distinguishable from random by this process; we then retrained the IQP model with the same hyperparameters, but with all-to-all two-qubit gates, which resulted in better results (we did not do this for the bitflip model due to its poor initial performance). For the RBM, the grid search selected a model with 250 hidden components, but the trained model suffered from extreme mode collapse, as can be seen from the fact that many of the diagonal elements of the covariance matrix are zero. For the EBM model, we attempted to build the graph structure into the energy function by masking and symmetrizing the layer weights so that the layers are equivalent to graph convolution layers with node feature dimension 1. The grid search selected a very simple linear neural network architecture with only 484 parameters, resulting in a covariance matrix that is indistinguishable from random (not shown here), although there is some improvement over random data for the test MMD loss. 

As a result, the IQP model was the only model that was able to produce reasonable results for this dataset. As before, the two body correlations shown in the covariance matrix mirror the overall structure of the true distribution, albeit with significantly weaker correlations. Surprisingly, for the smallest bandwidth (which probes correlations between 17 bits on average), the bitflip model achieves the best test MMD score, although it is unclear to use why this is the case. It was not possible to obtain KGEL plots for this experiment (the solver could not find solutions), likely because no model produced a distribution sufficiently close to the ground truth. 

\begin{figure*}
    \centering
    \includegraphics[width=0.8\textwidth]{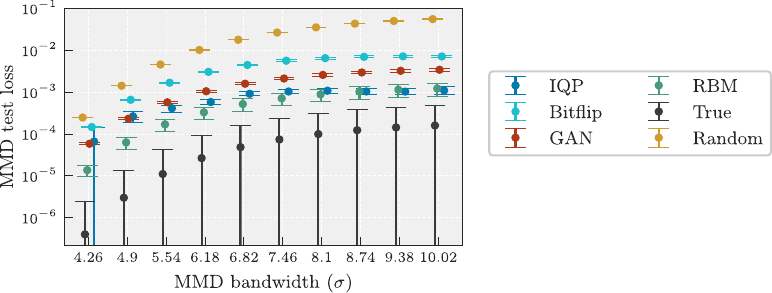}
    \caption{The test MMD values obtained for each of the trained models for the genomic dataset.}
    \label{fig:genomic}
\end{figure*}

\begin{figure*}
    \centering
    \includegraphics[width=1.0\textwidth]{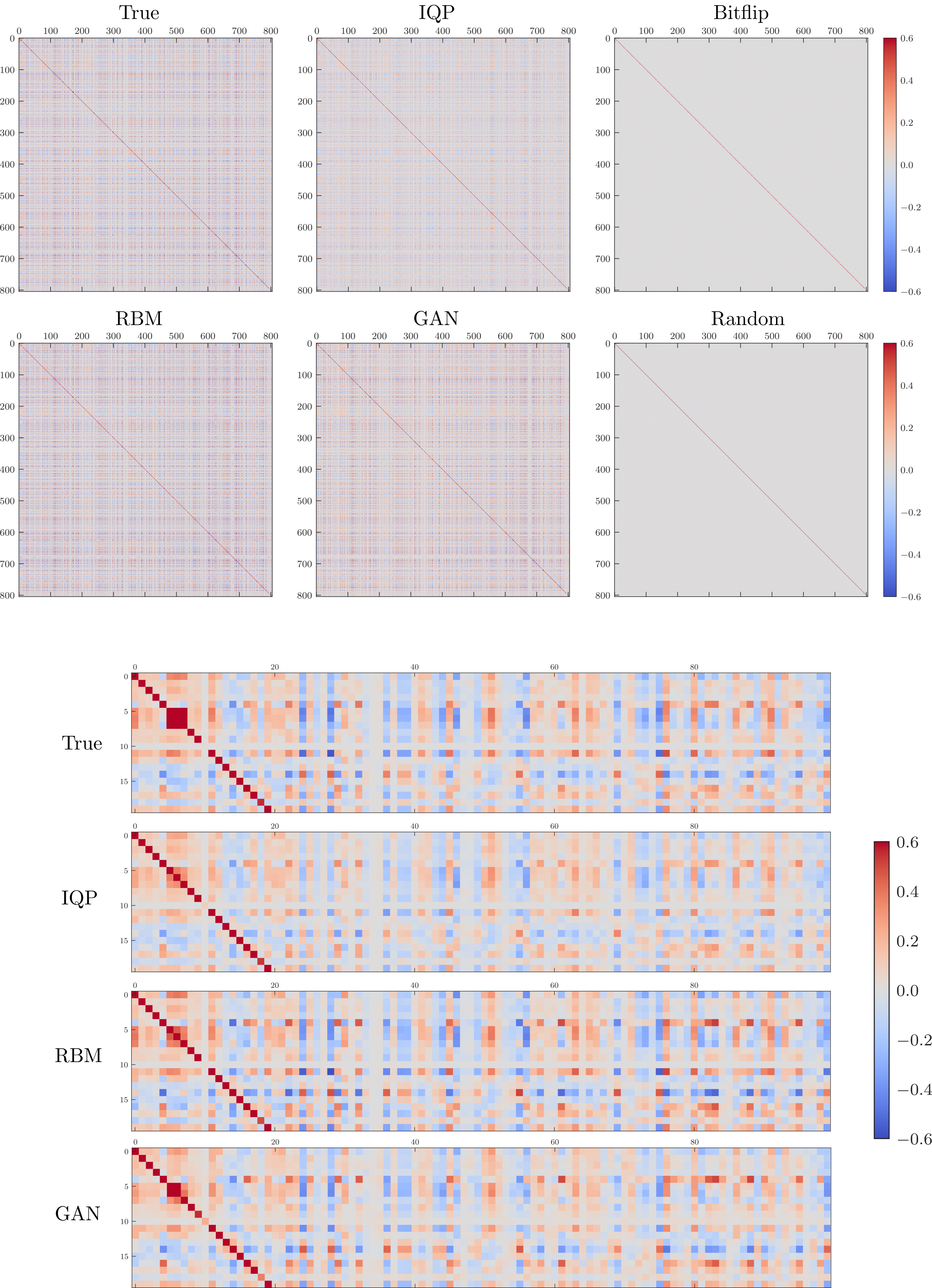}
    \caption{(top) Full covariance matrices for the six models for the genomic dataset. The RBM and GAN models were trained in \citet{yelmen2021creating}. (bottom) Zoom of the covariance matrices for the true, IQP, RBM and GAN models showing the covariances between the first 20 and 100 dimensions of the data. }
    \label{fig:genomic_cov}
\end{figure*}

\subsection*{Experiment 6: Genomic dataset}
This dataset was taken from the work of \citet{yelmen2021creating}, which trains classical generative models to learn the distribution of alleles at 805 highly differentiated biallelic single nucleotide polymorphisms (SNPs) of the human genome. Since the alleles take one of two values, this results in a distribution over bitstrings, with the presence of a 1 at a given location marking the presence of the variant allele. The dataset was constructed from genetic data from 2504 individuals from the human genome project, which results in a dataset of size 5008 (since each individual has two haplotypes) of bitstrings of length 805. We split this data into train and test sets with a test set ratio of 1/3. In \citet{yelmen2021creating}, the authors train an RBM model and a general adversarial network (GAN) model to learn the distribution, and provide samples from the trained models in a corresponding repository (\href{https://gitlab.inria.fr/ml_genetics/public/artificial_genomes}{gitlab.inria.fr/ml\_genetics/public/artificial\_genomes}). This allows us to compare our results to theirs via the corresponding MMD values and KGEL test. 

We trained an IQP and bitflip model with all-to-all connected two-qubit gates, resulting in models with 324415 parameters. The trained IQP model produced results that were competitive with the published results for the RBM and GAN, with the values for the test MMD$^2$ generally lying in between the two classical models. The covariance plots (Fig.~\ref{fig:genomic_cov}) show that the IQP model can capture the overall structure of the data, but again shows slightly weaker correlations than the ground truth. It is worth noting that the classical models in \citet{yelmen2021creating} appear to have be trained on the entire data (train plus test), which obviously gives these models an unfair advantage for this particular experiment. We also discovered that the samples from the RBM model are correlated (as may be expected when sampling the RBM from via Markov chain Monte Carlo methods), since shuffling the samples before estimating the mean and variance of the test MMDs resulted in very different scores. We therefore cannot fully trust the shown MMD$^2$ values for the RBM since we cannot guarantee i.i.d.\ samples as required by the estimator \eqref{eq:MMD_greton}. To produce the shown values, we first shuffled and batched the RBM samples before using \eqref{eq:MMD_greton} to estimate means and variances. The MCMC sampling of the RBM model was also seeded by additional genomic data, which was not available to us.

\section{Discussion of results: a glimmer of hope?}\label{sec:discussion}
Do the above results suggest that parameterised IQP circuits offer a fruitful path towards useful generative quantum machine learning? In the following we present a number of reflections with this question in mind. 

\subsection{Scalable training is a reality}
The quantum machine learning literature is full of doubts about the ability to scale models beyond the small scale numerical examples that appear in many works (\cite{mcclean2018barren,bittel2021training,cerezo2023does,anschuetz2022quantum,cerezo2021cost,rudolph2024trainability,sweke2021quantum}). Our results clearly demonstrate however that there exist scenarios in which it is possible to train large quantum circuits to non-trivial parameter configurations in a matter of hours. We therefore suspect that the training of large parameterized quantum circuits may be much easier in practice than is widely assumed, and that theoretical results that rest on broad and mathematically convenient assumptions should be taken with a healthy dose of skepticism regarding their practical implications. Without this we risk repeating history: the early literature on neural networks was also filled with similar concerns about the possibility of training large neural networks (\cite{bengio1994learning,blum1988training,judd1990neural,auer1995exponentially}), and the pessimistic predictions of \citet{minsky2017perceptrons} are believed by many to have been a significant contributor to the well-known AI winter of the 1980s. 

Regarding the specific scaling of our approach, it is likely possible to push far beyond the circuit sizes considered in this work. In particular, as we show in \citet{iqpopt}, the computational cost of estimating expectation values (to fixed precision) scales linearly with both the number of qubits and the number of circuit parameters. With the current version of the IQPopt package, it is possible to estimate the MMD loss (to the same precision considered in our experiments) for circuits with ten thousand qubits and one million gates in roughly one minute on a single CPU compute node. Improving the way the code deals with sparse objects and combining this with GPU-accelerated basic linear algebra subroutines like cuSparse, 
we expect to be able to go well beyond even these limits (see \citet{iqpopt} for more details). 


\subsection{The IQP model can compete with established classical models}
The IQP model achieved the best MMD test scores in three of the four experiments in which all models were trained. While for the 2D Ising dataset this was in part due to the fact that the IQP model was trained to minimize the MMD loss on the training data (unlike the EBM or RBM models), for the two large datasets (D-Wave and scale free), the IQP model clearly outperformed the classical models, which failed to produce convincing results. The reasons why the classical energy based models failed to train are still not fully clear. Our RBM model appears to have suffered from either mode imbalance or mode collapse in every experiment. This seems to have been much less of a problem for the IQP model, and it would be interesting to investigate further the behavior of the IQP model when learning highly multi-mode data. The failures of the EBM might have been due to insufficient sampling from the model during training and/or evaluation, since energy based models are known to be computational intensive to train. Even if this is the case, it suggests that parameterized IQP circuits have a training efficiency advantage, since both models used a similar amount of compute during training. We stress that although the results of the IQP circuit are generally better than the classical models, we are not claiming that better classical results could not be obtained with a wiser choice of hyperparameter grids or initialisation strategies. However, we believe that the ease in which we obtained superior results with the IQP model is a promising sign that warrants further attention. 

We chose an RBM and EBM as our classical comparisons due to the fact that they both work naturally on binary datasets, and because graph based correlation structures can be encoded naturally into both an EBM (through the energy function) and the IQP model (through the choice of gate set). Furthermore, the sci-kit learn implementation of the RBM is an established model that has been used extensively and therefore provides a basic sanity check. Nevertheless, neither model could be described as state-of-the-art. Certainly, there are other models that suit binary data and could serve as worthy comparisons, such as variational autoencoders (VAEs) with Bernoulli decoder networks, Markov random fields, diffusion models (with a Bernoulli diffusion process), generative adversarial networks and autoregressive models (such as transformers). There are aspects of the IQP model which may be advantageous with respect to each of these however. Unlike energy-based models (Markov random fields, diffusion models) and VAEs, which due to finite MCMC sampling or the use of the evidence lower bound (ELBO) both train on biased versions of the desired log likelihood function, the IQP model is trained on provably unbiased gradients of the MMD loss. Autoregressive models have a rigid sequential structure, which makes it unnatural to encode undirected graph correlation structures, as can be done naturally in gate based quantum models (\cite{bako2024problem}). Finally, GANs share many similarities with quantum models since they are also implicit models (and have also been trained via MMD loss functions, \citep{li2017mmd}), however the need to backpropagate continuous gradients through the network means that are not naturally suited to binary data, whereas the IQP model is natively binary. All of this suggests that structured, high dimensional binary datasets may be well suited to parameterized IQP models. It would be interesting to understand if and how the use of the Gaussian kernel can be adapted to other kernel choices, since the use of task dependent kernel functions is likely necessary to mitigate the curse of dimensionality and achieve genuine utility for problems of interest.

\subsection{Barren plateaus do not appear to be a problem}\label{sec:bps}
One of the biggest surprises from our experiments was how easy it was to train the IQP circuits, with convergence being reached smoothly in a small number of steps for all problems (Fig.~\ref{fig:lossplots}). 
Curiously, although our circuits were very large, we did not encounter problems related to barren plateaus (\cite{mcclean2018barren,ragone2024lie,arrasmith2022equivalence}). It is still unclear however whether this is due to our choice of parameter initialisation or because the anstaz is inherently free from barren plateaus. We remark that, even though the dynamical Lie algegras (DLAs) of our circuits are of size $\text{poly}(n)$ (since all gate generators commute), diagnostics from \citet{ragone2024lie} cannot be applied since the input state and observables are diagonal in the Z basis and are therefore not contained in the DLA.

Nevertheless, from \eqref{eq:expval_intefere} we can still see that for the observable $Z_1$ (a Pauli Z on the first qubit), the partial derivatives $\partial\langle Z_1\rangle/\partial{\theta_j}$ will concentrate exponentially around zero as the circuit size is increased. For example, considering circuits with all-to-all connected two-qubit gates, one finds that $\Omega$ in \eqref{eq:expval_intefere} is the empty set, since any gate generators that anticommute with $Z_1$ must act non-trivially on distinct qubits.  The expectation value is therefore $\langle Z_1\rangle=\prod_k \cos(2\theta_k)$, where the product is over all $k$ such that the corresponding generator acts non-trivially on the first qubit. One sees that under random initialisation the variance of $\langle Z_1\rangle$ 
\begin{align}
    \text{Var}[\langle Z_1\rangle] = \mathbb{E}_{\btheta}\left[\prod_k \cos^2(2\theta_k)\right] -  \mathbb{E}_{\btheta}\left[\prod_k \cos(2\theta_k)\right]^2
\end{align}
%
must decay exponentially to zero with $n$ since there are $n$ terms in each product. Initializing the $\theta_j$ uniformly at random therefore leads to exponential concentration of $\langle Z_{1}\rangle$, and therefore of the gradient too (\cite{arrasmith2022equivalence}). 

\begin{figure*}
\includegraphics[width=\textwidth]{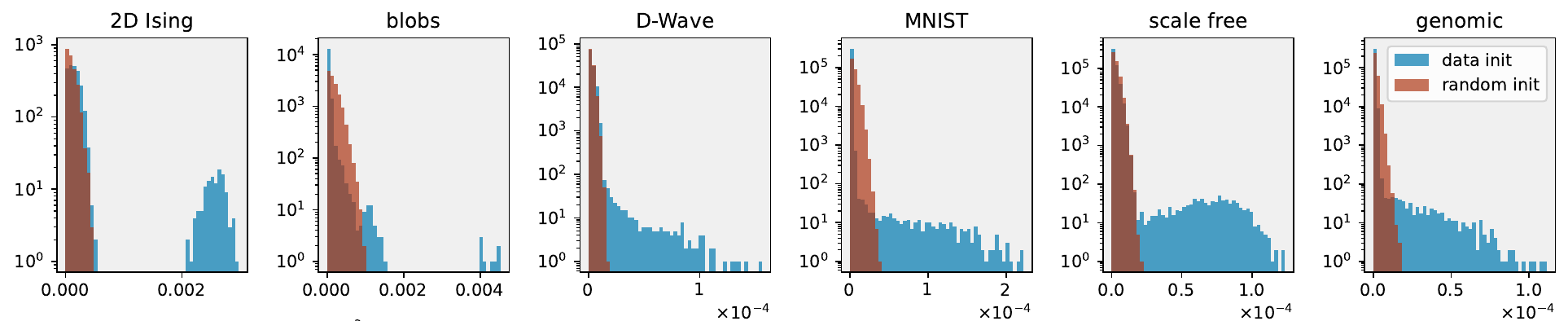}
\caption{\label{fig:grads}Histogram showing the distribution of the (estimated) magnitudes of gradient components for each of the datasets under either our data-dependent initialisation or a random initialisation (the y-axis is the number of components with magnitude in the given bin). For large datasets, we suspect that the variance of the distribution for the random initialisation is due to the error in the estimator \eqref{eq:mmd_beast} rather than the presence of large components.}
\end{figure*}

For higher order expectation values such as $\langle Z_1 Z_2\rangle$ the situation is not as clear. For gate sets with single qubit and all-to-all connected two-qubit gates we see that there are a total of $2n-2$ generators that anticommute with $Z_1 Z_2$, however the set $\Omega$ is now exponentially large since there are many ways to multiply these operators to the identity. The value of $\langle Z_1 Z_2\rangle$ is therefore not necessarily exponentially small, and a similar argument to above does not work. Since the MMD$^2$ is comprised of such expectation values, it is therefore possible that gradients do not concentrate exponentially for this loss. Note that for the bitflip model, concentration does occur however. Since for this model the expectation values are given simply by \eqref{eq:expval_bitflip},
\begin{align}
    \langle Z_{\a} \rangle = \prod_{j\vert \{X_{\g_j},Z_{\a}\}=0} \cos(2\theta_j)
\end{align}
exponential concentration of $\langle Z_{\a}\rangle$ will always occur by the same argument as above when initializing the parameters at random with all-to-all connected two-qubit gates. 

To investigate things further, we estimated the gradient vector via \eqref{eq:mmd_beast} for the IQP model for each of the six experiments, initializing the parameters either uniformly at random in $[0,2\pi]$ or via our data-dependent strategy (see Fig.~\ref{fig:grads}). Interestingly, we see that the data-dependent initialisation results in a small fraction of gradient components that have relatively large magnitudes, which is likely why we are able to train the model. We note that although \citet{rudolph2024trainability} suggest to train logarithmic depth circuits with a maximum bandwidth $\sigma=\sqrt{n/4}$ in order to avoid barren plateaus, we successfully train outside of this regime, since our gate sets have linear depth, and the bandwidths we use are significantly smaller than $\sqrt{n/4}$ (thus probing higher order correlations). Even though the random initialisation appears also to produce some relatively large gradient components, we suspect that this is an artifact of the finite precision of the estimator \eqref{eq:mmd_beast} rather than a lack of barren plateaus, and it may be that the true gradients are much smaller than this. Indeed, with further analysis we were unable to demonstrate that any of the `large' gradient components were different from zero in a statistically significant manner. Moreover, the fact that we failed to train large models with random initialisation also supports the idea that the true gradients are much smaller. 

It therefore appears that our data-dependent initialisation strategy was crucial to obtain good results, and may have successfully mitigated problems of barren plateaus. If this is the case, it suggests that much of the pessimism regarding the trainability of parameterized quantum circuits may be unfounded, and that the presence or absence of barren plateaus should not be used as a litmus test for whether or not a particular circuit structure is deemed trainable. Rather, we suspect that---as was the case for deep neural networks (\cite{glorot2010understanding,klambauer2017self})---novel parameter initialisations will be crucial to achieve good performance for large parameterized quantum circuits even if they exhibit barren plateaus under random initialisation (see \cite{patti2021entanglement,grant2019initialization,zhang2022escaping,kulshrestha2022beinit,sack2022avoiding} for some proposed solutions). We remark that we did not employ data-dependent initialisation strategies for the two energy-based models, and so part of the relative success of the IQP model may be due to this. It would be interesting to see whether data-dependent initialisations (for example, \cite{bereux2025fast}) would lead to significantly improved results on the classical side. In any case, we stress that the fact that a simple, scalable and effective data-dependent parameter initialisation exists for the IQP model is a strength that may not be generally be present in other models. 

\subsection{Coherence helps}
Another clear signal that can be seen from the results is that coherence appears crucial to obtain good results, since the bitflip model produced very poor results on all but the low dimensional datasets. Indeed, since the IQP and bitflip models had identical hyperparameter grids and initialisation and training strategies, the only thing differentiating these models is the form of the expectation values given by either \eqref{eq:expval_intefere} or \eqref{eq:expval_bitflip}. We therefore expect that additional terms appearing in the \eqref{eq:expval_intefere} are critical in either increasing expressivity of the model or improving the loss landscape, although we have not investigated this. An answer to this question would help understand not only the IQP model, but may give hints about differences between quantum and classical models more generally.  

\subsection{Do we need more expressivity?}
In all experiments the IQP models seem to have captured the general structure of the data without serious mode imbalance, but tend to produce weaker correlations than the true distributions, as can be seen from the covariance matrix plots. This suggests that stronger results might have be obtained with more expressivity. The most straightforward way to achieve this would be to add more gates to the circuits (for example, including some three or more qubit gates), however it is not clear how to do this for high dimensional datasets since the number of gates on more than two qubits becomes extremely large. Even this may not be enough however, since despite using all gate generators with weight of 6 or less (nearly fifteen thousand parameters), the IQP model for the binary blobs dataset still failed to learn the distribution well. This suggests that there may be more fundamental issues with the expressivity of the model class, which is supported by the lack of universality that occurs already for two-qubit circuits. We do expect however that the model class could be enlarged by extending the simulation algorithm to a larger class of states and/or measurements however, and a natural next step would be to attempt this in order to improve the performance on the blobs dataset, and ideally prove universality (in the sense of Sec.~\ref{sec:expressivity}). 

\subsection{Quantum models for quantum data?}
A commonly repeated mantra in the quantum machine learning community is that quantum models are best suited to `quantum data', that is, data that originates from an inherently quantum process. Interestingly, our results appear to support this, since the IQP model performed far better than any of the classical models on the only dataset (D-Wave) related to a quantum process. We stress however that this is just one experiment, and it is still unclear whether the D-Wave distribution is a genuinely difficult distribution for classical models to capture, or whether our classical models failed for other reasons. In our opinion it would not be surprising if further hyperparameter exploration or alternative classical models lead to significantly better classical results, and more investigation in this direction is needed before making any serious claims of quantum advantages. 

\section{Outlook: Should we prioritize scalable training?}\label{sec:outlook}
Building a successful machine learning model invariably requires a combination of two things: a suitable inductive bias that incorporates knowledge of the problem, and an architecture that scales to large parameter counts and dataset sizes. Both can be seen as remedies to the curse of dimensionality, in the former case by constraining the possible models to a smaller, more relevant space, and in the latter case by allowing for training with larger datasets and more expressive models. If the recent successes of deep learning have taught us anything it is that scaling might be the most critical of these two ingredients, with such an enormous push for scalability (\cite{bahri2024explaining}) that we might even be running out of internet data on which to train large language models (\cite{Ilyatalk}). 

Given this, it is extraordinary that---despite a significant drive to understand and encode biases in quantum models (\cite{larocca2022group,east2023all,chinzei2024trade,wakeham2024inference,wiersema2025geometric, meyer2023exploiting,bowles2023contextuality,kubler2021inductive,le2025symmetry,gili2024inductive,klus2021symmetric})---the vast majority of variational quantum machine learning works either ignore the issue of scalability, or equate it to the presence of barren plateaus.
%
In particular, little attention is given to what may be the most fundamental barrier to scalability in quantum machine learning: the fact that extracting gradient information from quantum circuits appears in general to be very costly (\cite{abbas2024quantum, bowles2023backpropagation}) and effectively forbids training most circuit structures at scale. 

We therefore hope that more effort will be directed to understanding how to build quantum models that can scale in practice. This does raise an interesting question however: since both scalability and useful inductive bias are needed for powerful quantum models, should we prioritize one over the other? In our opinion, since scalable models appear to be both necessary and rare, then it may be best to first understand what approaches can be scaled, and only then concentrate on how to encode biases into the corresponding models\footnote{As an example of the reverse approach, note that capsule networks (\cite{sabour2017dynamic}), widely regarded as having a superior inductive bias than convolutional neural networks, never took off because of the relative difficulty in scaling.}. This is the situation we are faced with in this work, and one central challenge now is to better understand what biases are either present or can be encoded into parameterized IQP circuits. At the same time, we imagine that we are only touching the surface of models that can be scaled in a similar way, and we hope our approach inspires other works that eventually culminate in both useful and scalable models. Ironically, it may be that generative learning---typically the most compute-hungry setting---provides the most promising route to achieve this since it opens the possibility of training on classical hardware alone.

\section{Acknowledgments}
We thank the authors of the package \href{https://github.com/johannesjmeyer/rsmf}{rsmf} that we used when creating figures. JB acknowledges useful comments and  
conversations with David Wakeham, Maria Schuld, Alexia Salavrakos and Patrick Huembeli. JB thanks Lee O'Riordan and Sanchit Bapat for help with HPC resources. This work has been supported by the Government of Spain (Severo Ochoa CEX2019-000910-S, FUNQIP and European Union NextGenerationEU PRTR-C17.I1), Fundació Cellex, Fundació Mir-Puig, Generalitat de Catalunya (CERCA program) and European Union (PASQuanS2.1, 101113690). ER is a fellow of Eurecat's \emph{Vicente López} PhD grant program.

\bibliographystyle{abbrvnat}
\bibliography{main}

\appendix

\section{Unbiased estimates of the MMD$^2$}\label{app:unbiased}
Here we prove the unbiasedness of the estimators that appear in the main text. We first cover the estimate \eqref{eq:MMD_greton}, which uses samples from both distributions, and then construct an unbiased estimator for the case where one distribution is a parameterised IQP circuit for which we do not have sample access. We recall from Def.~\eqref{def:mmd} the definition of the MMD$^2$ for distributions $p$ and $q$:
\begin{align}\label{defmmd}
   \text{MMD}^2(p, q) = \mathbb{E}_{\x,\y\sim p}\left[k(\x,\y)\right] - 2\mathbb{E}_{\x\sim p, \y\sim q}\left[k(\x,\y)\right] +  \mathbb{E}_{\x,\y\sim q}\left[k(\x,\y)\right].
\end{align}

\subsection{Case 1: samples are available from both distributions}

We start by showing why the expression \eqref{eq:MMD_greton}---which uses samples from $p$ and $q$---is an unbiased estimate of the MMD$^2$ (following \cite{JMLR:v13:gretton12a}). Suppose we have a set of samples $\mathcal{X}=(\x_1\cdots,\x_N)$ with $\x_i\sim p(\x)$ and  set of samples $\mathcal{Y}=(\y_1\cdots,\y_M)$ with $\y_i\sim q(\y)$. It is claimed that the expression
\begin{align}
    \hat{\text{MMD}}^2 = \frac{1}{N(N-1)} \sum_{i\neq j} k(\x_i, \x_j) - \frac{2}{NM}\sum_{i,j}k(\x_i,\y_j)+ \frac{1}{M(M-1)}\sum_{i\neq j}k(\y_i,\y_j)
\end{align}
of Eq.~\ref{eq:MMD_greton} is an unbiased estimator of MMD$^2(p,q)$. To see why this is the case, let us look at the first term 
\begin{align}\label{gretonistheman}
    \frac{1}{N(N-1)}\sum_{i\neq j} k(\x_i,\x_j).
\end{align}
Taking the expectation with respect to sampling the set $\mathcal{X}$ we find
\begin{align}
    \mathbb{E}_{\mathcal{X}}[\frac{1}{N(N-1)}\sum_{i\neq j} k(\x_i,\x_j)] =   \frac{1}{N(N-1)}\sum_{i\neq j}\mathbb{E}_{\mathcal{X}}[ k(\x_i,\x_j)] = \mathbb{E}_{\x,\y \sim p}[k(\x,\y)].
\end{align}
The last equality above follows because given an $\mathcal{X}$,  $\x_i$ and $\x_j$ are i.i.d.\ samples from $p$ if $i\neq j$, and so averaging a specific $ k(\x_i,\x_j)$ over sampling of the sets $\mathcal{X}$ is equivalent to averaging $k(\x,\y)$ with $\x$, $\y$ being sampled i.i.d.\ from $p$. We see it was necessary to remove the terms $k(\x_i,\x_i)$  from the sum since $\mathbb{E}_{\mathcal{X}}[ K(\x_i,\x_i)]= \mathbb{E}_{x\sim p}[k(\x,\x)] = 1$, which leads us to a denominator $N(N-1)$ rather than $N^2$. An analogous situation occurs for the last term, hence its analogous form. For the cross term 
\begin{align}
    \frac{2}{NM}\sum_{i,j}k(\x_i,\y_j)
\end{align}
We see that it is an unbiased estimator of $2\mathbb{E}_{\x\sim p \y\sim q}[k(\x,\y)]$ since 
\begin{align}
\mathbb{E}_{\mathcal{X},\mathcal{Y}}\left[\frac{2}{NM}\sum_{i,j}k(\x_i,\y_j) \right]= \frac{2}{NM}\sum_{i,j}\mathbb{E}_{\mathcal{X},\mathcal{Y}}[k(\x_i,\y_j)]&= \frac{2}{NM}\sum_{i,j}\mathbb{E}_{\x\sim p, \y\sim q}[k(\x,\y)] \\ &= 2\mathbb{E}_{\x\sim p, \y\sim q}[k(\x,\y)].
\end{align}
A similar situation to this happens when constructing an unbiased estimator for the parameterised IQP model, which we cover in the following section. 
 
\subsection{Case 2: One of the distributions is a parameterised IQP circuit}
We now construct the unbiased estimator that we use to train our models. From Prop.~\ref{prop:mmd} and \eqref{eq:expvalexp} we know that
\begin{align}\label{iqpterm}
    \mathbb{E}_{\x\sim p_1, \y\sim p_2}[k(\x,\y)] = \mathbb{E}_{\a\sim \mathcal{P}(\a)}\mathbb{E}_{\z_1\sim U}\big[f_{p_1}(\a,\z_1)\big]\mathbb{E}_{\z_2\sim U}\big[f_{p_2}(\a,\z_2)\big]
\end{align}
where the functions $f_{p_k}$, $k=1,2$ take the form
\begin{align}\label{eq:quantumfn}
   f_{p_k}(\a,\z_k) = \cos \sum_j \theta_j (-1)^{\g_j \cdot \z_k}(1-(-1)^{\g_j\cdot \a})
\end{align}
if $p_k$ corresponds to a parameterised IQP quantum model, and 
\begin{align}\label{classicalterm}
    f_{p_k}(\a,\z_k) = \mathbb{E}_{\x\sim p_k}\left[ (-1)^{\x\cdot \a }\right] = \langle Z_{\a} \rangle_{p_k}
\end{align}
if $p_k$ corresponds to a classical distribution from which we can sample. Note that there is no dependence on $\z_k$ in this case. We will work through the terms in \eqref{defmmd} one by one for clarity. 

\subsubsection{The last term}
We start with the last term in \eqref{defmmd}:
\begin{align}\label{firstterm}
    \mathbb{E}_{\x,\y\sim q_{\btheta}}[k(\x,\y)]. 
\end{align}
Our estimator for this term is
\begin{align}
    \frac{1}{|\mathcal{A}| |\mathcal{Z}|(|\mathcal{Z}|-1)}\sum_{i}\sum_{j}f_{q_{\btheta}}(\a_i,\z_j)\sum_{k\neq j}f_{q_{\btheta}}(\a_i,\z_k),
\end{align}
where $\{\a_i \sim \mathcal{P}_{\sigma}\}=\mathcal{A}$ is a batch of bitstrings specifying the observables and $\{\z_i \sim U\}=\mathcal{Z}$ a batch of uniformly sampled bitstrings. Taking the expectation value with respect to the sampling of sets $\mathcal{A}, \mathcal{Z}$ we find  
\begin{align}
    &\frac{1}{|\mathcal{A}| |\mathcal{Z}|(|\mathcal{Z}|-1)}\sum_{i}\sum_{j}\sum_{k\neq j} \mathbb{E}_{\mathcal{A},\mathcal{Z}}\big[ f_{q_{\btheta}}(\a_i,\z_j)f_{q_{\btheta}}(\a_i,\z_k) \big] \\
    =& \frac{1}{|\mathcal{A}| |\mathcal{Z}|(|\mathcal{Z}|-1)}\sum_{i}\sum_{j}\sum_{k\neq j} \mathbb{E}_{\mathcal{A}} \mathbb{E}_{\z_j\sim U}[f_q(\a_i,\z_j)]\mathbb{E}_{\z_k\sim U}[f_q(\a_i,\z_k)]  \\
    =& \frac{1}{|\mathcal{A}| |\mathcal{Z}|(|\mathcal{Z}|-1)}\sum_{i}\sum_{j}\sum_{k\neq j}\mathbb{E}_{{\a}\sim \mathcal{P}_{\sigma}(\a)} \mathbb{E}_{\z\sim U}[f_{q_{\btheta}}(\a,\z)]\mathbb{E}_{\z\sim U}[f_{q_{\btheta}}(\a,\z)] \\
    =& \mathbb{E}_{{\a}\sim \mathcal{P}_{\sigma}(\a)} \mathbb{E}_{\z\sim U}[f_{q_{\btheta}}(\a,\z)]\mathbb{E}_{\z\sim U}[f_{q_{\btheta}}(\a,\z)] 
\end{align}
since $\z_j$, $\z_k$ are i.i.d. samples for any $\mathcal{Z}$. From \eqref{iqpterm} it follows this is equal to the first term \eqref{firstterm}.

\subsubsection{The second term}
The second term 
\begin{align}\label{firstterm}
    \mathbb{E}_{\x\sim p, \y\sim {q_{\btheta}}}[k(\x,\y)]. 
\end{align}
does not pose any problems. We see from \eqref{iqpterm} that this is equal to 
\begin{align}\label{obvo}
    \mathbb{E}_{\x\sim p, \y\sim {q_{\btheta}}}[k(\x,\y)] = \mathbb{E}_{\a\sim \mathcal{P}_{\sigma}(\a)}\mathbb{E}_{\z\sim U}\big[f_{{q_{\btheta}}}(\a,\z)\big]\langle Z_{\a} \rangle_p
\end{align}
an empirical estimate from batches $\mathcal{A}$, $\mathcal{Z}$ and $\mathcal{X}=\{\x_k\sim p\}$ is
\begin{align}
     \frac{1}{|\mathcal{A}| |\mathcal{Z}| |\mathcal{X}|}\sum_{i}\sum_{j}\sum_kf_{q_{\btheta}}(\a_i,\z_j) (-1)^{\x_k\cdot \a_i}.
\end{align}
The expectation of this with respect to sampling batches $\mathcal{X}, \mathcal{A}, \mathcal{Z}$ is the same as \eqref{obvo}. Explicitly;
\begin{align}
    &\mathbb{E}_{\mathcal{X},\mathcal{A}, \mathcal{Z}} \left[\frac{1}{|\mathcal{A}| |\mathcal{Z}| |\mathcal{X}|}\sum_{i}\sum_{j}\sum_{k} f_{q_{\btheta}}(\a_i,\z_j) (-1)^{\x_k\cdot \a_i} \right] \\
    =& \mathbb{E}_{\mathcal{A}, \mathcal{Z}} \left[\frac{1}{|\mathcal{A}| |\mathcal{Z}|}\sum_{i}\sum_{j}f_{q_{\btheta}}(\a_i,\z_j)\mathbb{E}_{\mathcal{X}}\left[\frac{1}{|\mathcal{X}|}\sum_k (-1)^{\x_k\cdot \a_i}\right]\right]\\
     =&\frac{1}{|\mathcal{A}| |\mathcal{Z}|}\sum_{i}\sum_{j}\mathbb{E}_{\mathcal{A}, \mathcal{Z}}\left[f_{q_{\btheta}}(\a_i,
    \z_j)\langle Z_{\a_i} \rangle_p\right] =\mathbb{E}_{\a\sim \mathcal{P}_{\sigma}}\mathbb{E}_{\z\sim U}\left[f_{q_{\btheta}}(\a,
    \z)\langle Z_{\a} \rangle_p\right]
\end{align}
as required. 

\subsection{The first term}
For the first term we need to estimate $\mathbb{E}_{\x,\y\sim p}[k(\x,\y)]$ using a single set of samples $\mathcal{X}$ from $p$ corresponding to the training set. We have already seen how to do this in \eqref{gretonistheman}. An unbiased estimate is 
\begin{align}
    \frac{1}{|\mathcal{X}|(|\mathcal{X}|-1)}\sum_{i\neq j} k(\x_i,\x_j).
\end{align}
We can also construct an estimator that involves sampling over $\a$. From  \eqref{iqpterm} and \eqref{classicalterm} we have:
\begin{align}\label{idk}
   \mathbb{E}_{\x,\y\sim p}[k(\x,\y)] =  \mathbb{E}_{a\sim \mathcal{P}(\a)}\left[\langle Z_{\a} \rangle_p \langle Z_{\a} \rangle_p \right] =  \mathbb{E}_{\a\sim \mathcal{P}(\a)}\left[\mathbb{E}_{\x\sim p}[ (-1)^{\x\cdot \a }]\mathbb{E}_{\y\sim p}[  (-1)^{\y\cdot \a }]\right].
\end{align}
An unbiased estimator is
\begin{align}
    \frac{1}{|\mathcal{A}| |\mathcal{X}|(|\mathcal{X}|-1)}\sum_i\sum_j\sum_{k\neq j}(-1)^{\x_j\cdot \a_i }(-1)^{\x_k\cdot \a_i }.
\end{align}
Taking the expectation with respect to sampling $\mathcal{A}$ and $\mathcal{X}$ we find
\begin{align}
   &\mathbb{E}_{\mathcal{A},\mathcal{X}} \left[\frac{1}{|\mathcal{A}| |\mathcal{X}|(|\mathcal{X}|-1)}\sum_i\sum_j\sum_{k\neq j}(-1)^{\x_j\cdot \a_i }(-1)^{\x_k\cdot \a_i }\right]\\=& 
   \frac{1}{|\mathcal{A}| |\mathcal{X}|(|\mathcal{X}|-1)}\sum_i\sum_j\sum_{k\neq j}\mathbb{E}_{\mathcal{A},\mathcal{X}}[(-1)^{\x_j\cdot \a_i }(-1)^{\x_k\cdot \a }] \\=& 
   \frac{1}{|\mathcal{A}| |\mathcal{X}|(|\mathcal{X}|-1)}\sum_i\sum_j\sum_{k\neq j}\mathbb{E}_{\mathcal{A}}[\langle Z_{\a_i} \rangle_{p}\langle Z_{\a_i} \rangle_{p}] =  \mathbb{E}_{\a\sim \mathcal{P}_{\sigma}}[\langle Z_{\a} \rangle_{p}\langle Z_{\a} \rangle_{p}]
\end{align}
as required. 

\subsection{The full expression}
Putting all this together we arrive at the full estimator for the MMD$^2$:

\begin{multline}\text{MMD}_{\text{u}}^2(\mathcal{A}, \mathcal{Z}, \mathcal{X}, \btheta) = \frac{1}{|\mathcal{A}| |\mathcal{Z}|(|\mathcal{Z}|-1)}\sum_{i}\sum_{j}f_{q_{\btheta}}(\a_i,\z_j)\sum_{k\neq j}f_{q_{\btheta}}(\a_i,\z_k) \\ - \frac{2}{|\mathcal{A}| |\mathcal{Z}|  |\mathcal{X}|}\sum_{i}\sum_{j}\sum_{k}f_{q_{\btheta}}(\a_i,\z_j)(-1)^{\x_k\cdot \a_i} \\ + \frac{1}{|\mathcal{A}| |\mathcal{X}|( |\mathcal{X}|-1)}\sum_i\sum_j\sum_{k\neq j}(-1)^{\x_j\cdot \a_i}(-1)^{\x_k\cdot \a_i} 
\end{multline}
where $f_{q_{\btheta}}(\a,\z) = \cos \sum_j \theta_j (-1)^{\g_j \cdot \z}(1-(-1)^{\g_j\cdot \a})$. From the above sections it follows that
\begin{align}
\mathbb{E}_{\mathcal{A},\mathcal{Z},\mathcal{X}}\left[\text{MMD}_{\text{u}}^2(\mathcal{A}, \mathcal{Z}, \mathcal{X}, \btheta)\right] = \text{MMD}^2(p,q_{\btheta})
\end{align}
which completes the proof.

\section{A toy example exploiting coherence}\label{app:toy}

\begin{figure}
    \centering
    \includegraphics{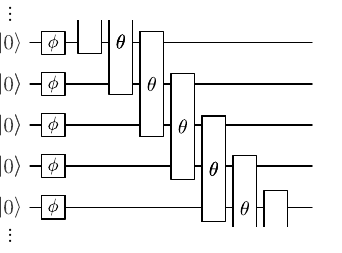}
    \caption{The circuit used for the toy example.}
    \label{fig:toy_circuit}
\end{figure}
Here we give a concrete example where interference is beneficial to prepare particular distributions. The aim is to give some intuition as to how coherence can be beneficial, but much more work is necessary to have a clearer understanding. 

We consider a parameterised IQP circuit of the form in figure \ref{fig:toy_circuit} where each qubit has a single qubit gate, and every triple of adjacent qubits share a gate (and we have periodic boundary conditions). Our aim will be to prepare a distribution with $\langle Z_i \rangle =0$ (that is, locally random), while aiming to maximize the values of $\langle Z_{i-1}\otimes Z_i \otimes Z_{i+1}\rangle \equiv \langle Z_{(i-1,i,i+1)} \rangle$. To keep things simple, we consider a translationally invariant model and distribution, with the same parameter $\phi$ for each single qubit gate and same parameter $\theta$ for each three qubit gate (see Fig.~\ref{ig:toy_circuit}). We first consider the condition $\langle Z_i \rangle=0$. Considering all generators that anticommute with $Z_i$ and using \eqref{eq:expval_intefere} we can see that $\Omega$ is the empty set so
\begin{align}
    \langle Z_i \rangle = \cos(2\phi)\cos^3(2\theta)
\end{align}
for both IQP and bitflip models. For $\langle Z_{i-1}Z_iZ_{i+1}\rangle$, $\Omega$ is not empty however since we have 
\begin{align}
    Z_{i-1} Z_i  Z_{i+1} Z_{(i-1,i,i+1)} = \mathbb{I}
\end{align}
and each single qubit operator anti-commutes with $Z_{(i-1,i,i+1)}$. The expectation value for the IQP circuit is therefore 
\begin{align}
    \langle Z_{(i-1,i,i+1)} \rangle & = \cos^3(2\phi)\cos^3(2\theta)+\cos^2(2\theta)\sin(2\theta)\sin^3(2\phi) \\
    & = \cos^2(2\theta)[\langle Z_i \rangle + \sin(2\theta)\sin^3(2\phi)]. \label{eq:interf}
\end{align}
To maximize this we set $\phi=\pi/4$ (so $\langle Z_i\rangle=0$ as required), and maximize $\cos^2(2\theta)\sin(2\theta)$. We find a maximum $2/(3\sqrt{3})\approx 0.385$ for the value
\begin{align}
    \theta = \arctan\sqrt{5-2\sqrt{6}}\approx0.308. 
\end{align}
If we consider a stochastic bitflip circuit with the same structure, we see from \eqref{eq:expval_bitflip} that we must have $\langle Z_{(i-1,i,i+1)}\rangle=0$ since the second term in \eqref{eq:interf} does not contribute and we require $\langle Z_i\rangle=0$. Thus, for this type of distribution we have an advantage relative to classically flipping the same bits, which is likely advantageous when learning from data with a similar bias.

\section{Non-universality for $n$ qubit circuits}\label{app:universality}
Here we show that the model class is not universal if the number of qubits is equal to the number of bits in the set of distributions we aim to parameterise. This follows from a simple argument for the case $n=2$. 

For two qubits, there are only three gates given by the generators $X_1$, $X_2$ and $X_{1}X_{2}$ (with parameters $\theta_1$, $\theta_2$, $\theta_{12}$). Since $\exp(i\theta G)=\cos(\theta)\mathbb{I} + i\sin(\theta)G$ for generators satisfying $G^2=\mathbb{I}$ (as in our case), we have that 

\begin{multline}
    \vert\bra{\vec{x}}U(\btheta)\ket{0}\vert^2  
     = \vert\bra{\vec{x}}e^{i\theta_{12}X_1X_2}(\cos\theta_1\ket{0}+i\sin\theta_1\ket{1})(\cos\theta_2\ket{0}+i\sin\theta_2\ket{1})\vert^2 \\ = 
    \vert\bra{\vec{x}}\bigg[\cos\theta_1\cos\theta_2(\cos\theta_{12}\ket{00}+i\sin\theta_{12}\ket{11}) -\sin\theta_1\sin\theta_2(\cos\theta_{12}\ket{11}+i\sin\theta_{12}\ket{00})  \\
    +i\cos\theta_1\sin\theta_2(\cos\theta_{12}\ket{01}+i\sin\theta_{12}\ket{10}) +i\sin\theta_1\cos\theta_2(\cos\theta_{12}\ket{10}+i\sin\theta_{12}\ket{01})\bigg]\vert^2.
\end{multline}

From this we see that 
\begin{align}
    p(00) = q_1q_2q_{12} + (1-q_1)(1-q_2)(1-q_{12}) \\
    p(10) = q_1(1-q_2)(1-q_{12}) + (1-q_1)q_2q_{12} \\
    p(01) = (1-q_1)q_2(1-q_{12}) + q_1(1-q_2)q_{12} \\
    p(11) = (1-q_1)(1-q_2)q_{12} + q_1q_2(1-q_{12})
\end{align}
where $q_1$, $q_2$, $q_{12}$ are probabilities given by 
\begin{align}
    q_1 = \cos^2\theta_1, \quad q_2 = \cos^2\theta_2, \quad q_{12}=\cos^2\theta_{12}.
\end{align}
Note that these are the exact same probabilities we would arrive at if we replace the quantum circuit by a classical stochastic circuit where we flip subsets of bits with probabilities $q_1$, $q_2$, $q_{12}$. For two qubits, the circuit therefore does not make use of interference to go beyond the classical limits of expressivity. Mathematically, this is due to the fact that branches of the wavefunction that result in the same outcome differ by a phase factor of $i$, which prevents interference due to these components being squared independently when computing probabilities. For larger numbers of qubits interference is possible between branches with the same phase, however. 

It remains to show that there are distributions that cannot be written in the above form. Consider the distribution 
\begin{align}
    (p(00), p(01), p(10), p(11)) = (\frac{1}{3},\frac{1}{3}, \frac{1}{3},0). 
\end{align}
With a little thought one sees that this can't be achieved by classically flipping subsets of bits independently, and thus also cannot be done with the IQP circuit, but we will prove it for completeness. The condition $p(11)=0$ implies 
\begin{align}
    (1-q_1)(1-q_2)q_{12} = - q_1q_2(1-q_{12}).
\end{align}
We see that we must ensure the right hand size is zero or a contradiction will be reached due to positivity of probabilities. We have three possibilities:
\begin{enumerate}
    \item $q_1 = 0 \implies q_2=1 \text{ or } q_{12}=0$
    \item $q_2 = 0 \implies q_1=1 \text{ or } q_{12}=0$
    \item $q_{12}=1 \implies q_1=1 \text{ or } q_2=1$
\end{enumerate}
Considering 1, we see that $q_1=0, q_2=1$ implies $p(00)=0$ and $q_1=0, q_{12}=0$ implies $p(10)=0$, so these are not valid solutions. A similar line of argument works for 2. Taking 3, we find
\begin{align}
    p(00) = q_0q_1 = \frac{1}{3} 
\end{align}
and so either $(q_0,q_1)=(1,\frac{1}{3})$ or $(q_0,q_1)=(\frac{1}{3},1)$. Finally we have 
\begin{align}
    p(10) = (1-q_1)q_2 
\end{align}
which must be equal to either $0$ or $\frac{2}{3}$ thus leading to a contradiction with the condition $p(10)=1/3$.

\section{Approximation of the KGEL} \label{app:kgel}
The formula that we will use to calculate this metric comes from \cite{ravuri2023understandingdeepgenerativemodels} (Eq.~6), which we repeat here:
\begin{align}\label{eq:kgel_app}
    \text{\normalfont{KGEL}}(\mathcal{X}_{\text{test}}, q_{\btheta}) = 
    \min_{\pi_i\}} D_{KL}(P_{\boldsymbol{\pi}}\vert\vert P_{\mathcal{X}_{\text{test}}}) \\[3pt] \mbox{\normalfont{subject to}} \quad
    \sum_{i=1}^n \pi_i 
    {\scriptscriptstyle \begin{bmatrix}
    k(\x_i, \t_1)\\
    \vdots \\
    k(\x_i, \t_W)
    \end{bmatrix}}
    =
    \mathbb{E}_{\y\sim q_{\btheta}} \begin{bmatrix}
    k(\y, \t_1)\\
    \vdots \\
    k(\y, \t_W)
    \end{bmatrix}. 
\end{align}
Since we don't have access to samples for the IQP circuit, we need another method to estimate the expectation value on the right hand side of the equation. Note that we have
\begin{align}
    \mathbb{E}_{y\sim q_{\btheta}}[k(\y, \t)] &= \mathbb{E}_{y\sim q_{\btheta}}[k(\y, \t)]\\
    &= \sum_{\y}q_{\btheta}(\y) k(\y, \t) \\
    &= \sum_{\y} \text{tr}[\ket{\y}\bra{\y}\rho_{\btheta}] k(\y, \t) \\
    &= \text{tr}[\sum_{\y} \ket{\y}\bra{\y} k(\y, \t) \rho_{\btheta}]  \\
    &= \text{tr}[O_{\text{KGEL}}(\t)\rho_{\btheta}]
\end{align}
Next, we write the observable $O_{\text{KGEL}}(\t)$ in terms of Pauli Z observables (taking inspiration from \cite{rudolph2024trainability}):
\begin{align}
    O_{\text{KGEL}}(\t) &= \sum_{\y} \ket{\y}\bra{\y} k(\y, \t) 
    = \sum_{\y} \ket{\y}\bra{\y} e^{\frac{-\sum_{i} (y_i-t_i)^2}{2\sigma^2}} \\
    &= \sum_{\y} \bigotimes_{i=1}^n\left[\ket{y_i}\bra{y_i} e^{\frac{-(y_i-t_i)^2}{2\sigma^2}}\right] \\
    &= \bigotimes_{i=1}^n \sum_{y_i=0,1} \left[\ket{y_i}\bra{y_i} e^{\frac{-(y_i-t_i)^2}{2\sigma^2}}\right] \\
    &= \bigotimes_{i=1}^n \left[\ket{0}\bra{0} e^{\frac{-t_i^2}{2\sigma^2}} + \ket{1}\bra{1} e^{\frac{-(1-t_i)^2}{2\sigma^2}}\right] \\
    &= \bigotimes_{i=1}^n \left[\frac{\mathbb{I} + Z}{2} e^{\frac{-t_i}{2\sigma^2}} + \frac{\mathbb{I} - Z}{2} e^{\frac{t_i-1}{2\sigma^2}}\right] \\
    &= \bigotimes_{i=1}^n \left[\frac{e^{\frac{-t_i}{2\sigma^2}} + e^{\frac{t_i-1}{2\sigma^2}}}{2} \mathbb{I} + \frac{e^{\frac{-t_i}{2\sigma^2}} - e^{\frac{t_i-1}{2\sigma^2}}}{2} Z\right] \\
    &= \bigotimes_{i=1}^n \left[\frac{1 + e^{\frac{-1}{2\sigma^2}}}{2} \mathbb{I} + (-1)^{t_i}\frac{1 - e^{\frac{-1}{2\sigma^2}}}{2} Z\right] \\
    &= \bigotimes_{i=1}^n \left[(1-p_{\sigma}) \mathbb{I} + (-1)^{t_i}p_{\sigma} Z\right] \\
    &= \sum_{\a \in\{0,1\}^n} (1-p_\sigma)^{n-|\a|}p_\sigma^{|\a|}(-1)^{\a \cdot\t} Z_{\a}.
\end{align}
Note that in several lines we have used the fact that $t_i\in\{0,1\}$. This means that
\begin{align}
    \mathbb{E}_{y\sim q}[k(\y, \t)] &= \text{tr}[O_{\text{KGEL}}(\t)\rho_{\btheta}]\\
    &= \sum_{\a \in \{0,1\}^n} (1-p_\sigma)^{n-|\a|}p_\sigma^{|\a|}(-1)^{\a \cdot\t} \langle Z_{\a}\rangle \\
    &= \sum_{\a \in \{0,1\}^n} \mathcal{P}_{\sigma}(\a)(-1)^{\a \cdot\t} \langle Z_{\a}\rangle , \\
    &= \mathbb{E}_{\a\sim\mathcal{P}_{\sigma}}[(-1)^{\a \cdot\t} \langle Z_{\a}\rangle].
\end{align}
Since this is an expectation with repsect to a bounded random variable, inverse polynomial additive errors can be obtained by sampling a batch $\vert \mathcal{A}\vert$ and computing an empirical mean, as we did for the the MMD loss. We can therefore estimate the vector on the right hand side of \eqref{eq:kgel_app} to the same error. 

\end{document}